\documentclass[traditabstract]{aa}

\ifx\pdfoutput\undefined
\usepackage{graphicx}
\else
\usepackage[pdftex]{graphicx}
\usepackage[pdftex,colorlinks,citecolor=blue,linkcolor=blue,pdfauthor= Matteo~Maturi,pdftitle =Filament~detection,pdfsubject=Cosmology]{hyperref}
\usepackage{epstopdf}
\fi

\usepackage{txfonts}
\usepackage{natbib}
\usepackage{hyperref}
\usepackage{rotating}
\usepackage{enumerate}
\usepackage{url}

\renewcommand{\d}{\mathrm{d}}

\usepackage{color}

\begin{document} 

\title{Multi-colour detection of gravitational arcs}


\author{Matteo Maturi\inst{1}\thanks{\email{maturi@uni-heidelberg.de}}
    \and Sebastian Mizera\inst{2}
    \and Gregor Seidel\inst{3}}

\titlerunning{}
\authorrunning{Maturi et al.}

\institute{
  $^1$~Zentrum f\"ur Astronomie der Universit\"at Heidelberg, Institut
  f\"ur Theoretische Astrophysik, Philosophenweg 12, 69120
  Heidelberg, Germany\\
  $^2$~University of Cambridge, The Old Schools, Trinity Lane, Cambridge CB2 1TN, United Kingdom\\
  $^3$~Max-Planck-Institut f\"ur Astronomie, K\"onigstuhl 17, 69117
  Heidelberg, Germany}

\date{\emph{Astronomy \& Astrophysics, submitted}}

\abstract{Strong gravitational lensing provides fundamental insights
  into the understanding of the dark matter distribution in massive
  galaxies, galaxy clusters and the background cosmology. Despite
  their importance, the number of gravitational arcs discovered so far
  is small. The urge for more complete, large samples and unbiased
  methods of selecting candidates is rising. A number of methods for
  the automatic detection of arcs have been proposed in the
  literature, but large amounts of spurious detections retrieved by
  these methods forces observers to visually inspect thousands of
  candidates per square degree in order to clean the samples. This
  approach is largely subjective and requires a huge amount of
  eye-ball checking, especially considering the actual and upcoming
  wide field surveys, which will cover thousands of square degrees.

  In this paper we study the statistical properties of colours of
  gravitational arcs detected in the 37 $\mathrm{deg}^2$ of the CARS
  survey.  We have found that most of them lie in a relatively small
  region of the $(g^\prime-r^\prime,r^\prime-i^\prime)$ colour-colour
  diagram. { To explain this property, we provide a model which
    includes the lensing optical depth expected in a $\Lambda$CDM
    cosmology that, in combination with the sources' redshift }
  distribution of a given survey, in our case CARS, peaks for sources
  at redshift $z\sim 1$. By further modelling the colours derived from
  the SED of the galaxies dominating the population at that redshift,
  the model well reproduces the observed colours.

  By taking advantage of the colour selection suggested by both data
  and model, we show that this multi-band filtering returns a sample
  83\% complete and a contamination reduced by a factor of $\sim6.5$
  with respect to the single-band arcfinder sample. New arc candidates
  are also proposed.

}

\maketitle

\section{Introduction}

Strong gravitational lensing is a powerful tool used to probe dark
matter and cosmology. The power is in its high non-linearity and
sensitivity to mass distributions and the geometry of space-time
\citep[e.g. see
  reviews,][]{1998LRR.....1...12W,1999grle.book.....S,2006glsw.conf...91K,bartelmann10}. It
finds many applications, such as (1) study of high redshift objects
which strongly magnify images allowing them to be examined further,
which would be very difficult otherwise
\citep[e.g.][]{2004ApJ...607..697K,2008ApJ...685..705R,2009ApJ...703L.132Z,2011MNRAS.414L..31R,2012ApJ...755L...7B,2013ApJ...762...32C},
(2) determination of the Hubble constant via time delays
\citep[e.g.][]{2008ApJ...679...17C,2010ApJ...711..201S,2012arXiv1208.6009T},
(3) measurement of dark matter amount and distribution of the lenses,
in particular of galaxy clusters
\citep{2005ApJ...621...53B,2009MNRAS.396.1985Z,2009A&A...500..681M},
and (4) study of intrinsic properties of dark matter, such as its
collisional cross section, thanks to major merger events between
galaxy clusters \citep{2004ApJ...604..596C, 2011MNRAS.417..333M}.
Moreover, giant arcs are a valuable tool to constrain cosmology, since
their number is very sensitive to cosmological parameters and
structure formation \citep{meneghetti13}. Of particular interest is
the tension between their predicted and observed number in the sky as
highlighted by
\cite{1995A&A...297....1B,1998A&A...330....1B,2000MNRAS.314..338M,2004ApJ...609...50D,2003MNRAS.340..105M},
even if recent studies alleviated this tension by introducing more
accurate cluster models based on numerical N-body simulations
\citep{2000MNRAS.314..338M,2004MNRAS.349..476T,2003MNRAS.340..105M,2005ApJ...635..795L,2005ApJ...633..768H}
and halo-models \citep{2006A&A...447..419F}.

So far, a relatively small number of gravitational arcs has been
found. In particular, only few giant arcs caused by galaxy clusters
are known. The search for them is mostly focused on bright X-ray
clusters, due to their large efficiency in producing strong lensing
features. This selection biases the actual sample, since these
clusters tend to be non-relaxed, limiting the possibilities of this
powerful observable. Up to now, only few hundred of cases have been
confirmed \citep[see
  e.g.][]{2004ApJ...600L.155F,2007A&A...461..813C,2008ApJS..176...19F,2008MNRAS.389.1311J,limousin09,verdugo11,2012ApJ...749...38M,2011ApJS..193....8B,2012MNRAS.420.3213O}. To
face these issues, automatic arc detection methods have been proposed
in the literature in order to produce unbiased samples, but the actual
methods still suffer from strong contamination and require heavy human
intervention by eye-ball checking thousands of candidates per square
degree. The need of a large amount of human resources and the
subjective outcome both limit the applicability of these procedures
\citep[see e.g.][]{2007A&A...461..813C,2012ApJ...749...38M}.

In this work, we aim at enlarging the sample of gravitational arcs by
using the arcfinder proposed by \citet{2007A&A...472..341S} combined
with a new colour selection procedure discussed in this work. The
method allows a strong reduction of the sample contamination
alleviating the efforts devoted to the final validation of the most
promising candidates. In this work we process the
CFHTLS-Archive-Research Survey (CARS, Erben et al., 2009)
data, covering 37 square degrees in order to verify the efficiency of
the method and to detect new gravitational arcs.

The paper is organised as follows: in Section~(2) and (3) we introduce
the basics of strong gravitational lensing and the data set
characteristics, in Section~(4) the arcfinder and its applications are
described, while in Section~(5) we characterize the colour properties
of arcs to be used for the subsequent selection. In Section~(6) the
full method is discussed and in Section~(7) we present the final
sample of arcs. Our conclusions are finally drawn in Section~(8).

\section{Basics of strong gravitational lensing}

All lensing quantities derive from the gravitational potential,
$\Phi$, of the matter placed along the line of sight on a thin plane
placed between the observer and the background sources
\begin{equation}
  \psi(\vec{\theta}) \equiv \frac{2}{c^2}
     \frac{D_{\rm ds}}{D_{\rm d}D_{\rm s}}
     \int 
     \Phi(D_{\rm d}\vec{\theta}, z)\,\d z \,.
\end{equation}
Here, $\psi(\vec\theta)$ is the so-called {\it lensing potential}
which depends on the angular position, $\vec\theta$, in the plane of
the sky, $c$ is the speed of light, $D_{\rm ds}$, $D_{\rm d}$ and
$D_{\rm s}$ are the lens-source, the observer-lens and the
observer-source angular-diameter distances, respectively.

Gravitational lensing maps the lens plane into the source plane via
the lens equation
\begin{equation}
  \vec\beta=\vec\theta-\nabla{\psi}(\vec\theta) \;,
\end{equation}
which can be linearised because sources, such as distant galaxies, are
much smaller with respect to the typical scale on which the lens
properties vary. The induced image distortion is thus expressed by the
Jacobian of the linearised lens equation
\begin{equation}
  A\equiv
  \frac{\partial\vec\beta}{\partial\vec\theta}=
  \left(\delta_{ij}-
  \frac{\partial^2\psi(\vec\theta)}{\partial\theta_i\partial\theta_j}
  \right)=\left(
  \begin{array}{cc}
    1-\kappa-\gamma_1 & -\gamma_2 \\
    -\gamma_2 & 1-\kappa+\gamma_1 \\
  \end{array}\right) \,,
  \label{eqn:jacobian}
\end{equation} 
where $\kappa$ is the convergence and $\gamma=\gamma_1+i\gamma_2$ is
the complex shear. Since $A$ is symmetric, it can always be
diagonalized and its two real eigenvalues, $\lambda_t=1-\kappa-\gamma$
and $\lambda_r=1-\kappa+\gamma$, represent the distortion of an
infinitesimal source in tangential and radial directions relative to
the lens centre, respectively. The length-to-width ratio ($L/W$
hereafter) of the image is thus defined as $q=\lambda_t/\lambda_r$ and
its magnification factor reads
\begin{equation}\label{eq:magnification}
  \mu=\frac{1}{(1-\kappa)^2-\gamma_1^2-\gamma_2^2} \;.
\end{equation}
For more details on gravitational lensing see for example
\cite{1998LRR.....1...12W}, \cite{1999grle.book.....S} or
\cite{bartelmann10}.

\begin{figure}[!t]
  \centering
  \includegraphics[width=0.49\hsize]{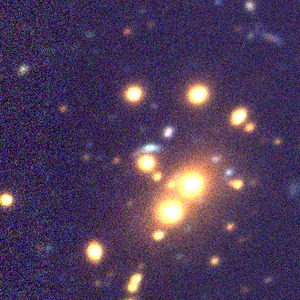}
  \includegraphics[width=0.49\hsize]{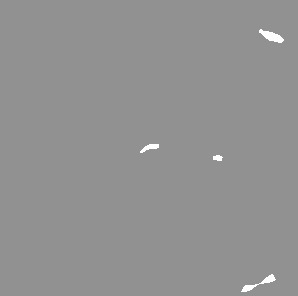}
  \caption{One of the arcs detected by \cite{2012ApJ...749...38M}, in
    the centre, used to initialize arcfinder parameters. The arc is
    clearly detected and the number of false detections is very small,
    in comparison to all objects in the field, already at the
    segmentation stage, before further selection process.}
  \label{fig:arcsegment}
\end{figure}

\section{Data set}\label{sec:dataset}

To test our arcfinding method and possibly discover new strong lensing
features, we processed stacked images belonging to the
CFHTLS-Archive-Research Survey (CARS) \citep{2009A&A...493.1197E}, a
set of three high-galactic-latitude patches covering a total of 37
$deg^2$ and produced with the publicly available observations obtained
with the MegaPrime camera mounted an the Canada-France-Hawaii
Telescope (CFHT) within the Canada-France-Hawaii-Telescope Legacy
Survey\footnote{\url{www.cfht.hawaii.edu/Science/CFHLS}}. The
MegaPrime is an optical camera with a mosaic of $9 \time 4$ CCDs of
$2048 \time 4096$ pixels, each sampling 0.186 arcseconds over a field of
view of one square degree \citep[see e.g.][]{2003SPIE.4841...72B}. All
observations were obtained through the filters $u^\star$, $g^\prime$,
$r^\prime$, $i^\prime$ and $z^\prime$. The three patches of sky
covered by CARS are:

\begin{center}
  \begin{tabular}[!t]{cccc}
    field & RA & Dec. & area ($deg^2$)\\
    \hline
    W1 & 02:18:00 & −07:00:00 & 21\\
    W3 & 14:17:54 & +54:30:31 & 5 \\
    W4 & 22:13:18 & +01:19:00 & 11\\
  \end{tabular}
\end{center}
The deep co-added images were produced using the GaBoDS/THELI pipeline
on a pointing/colour basis after rejecting all exposures with a
problematic CFHT quality assessment \citep{2005AN....326..432E}. All
details about CARS can be found in \cite{2009A&A...493.1197E}.

\begin{figure*}[!t]
  \centering
  \includegraphics[angle=-90,width=0.32\hsize]{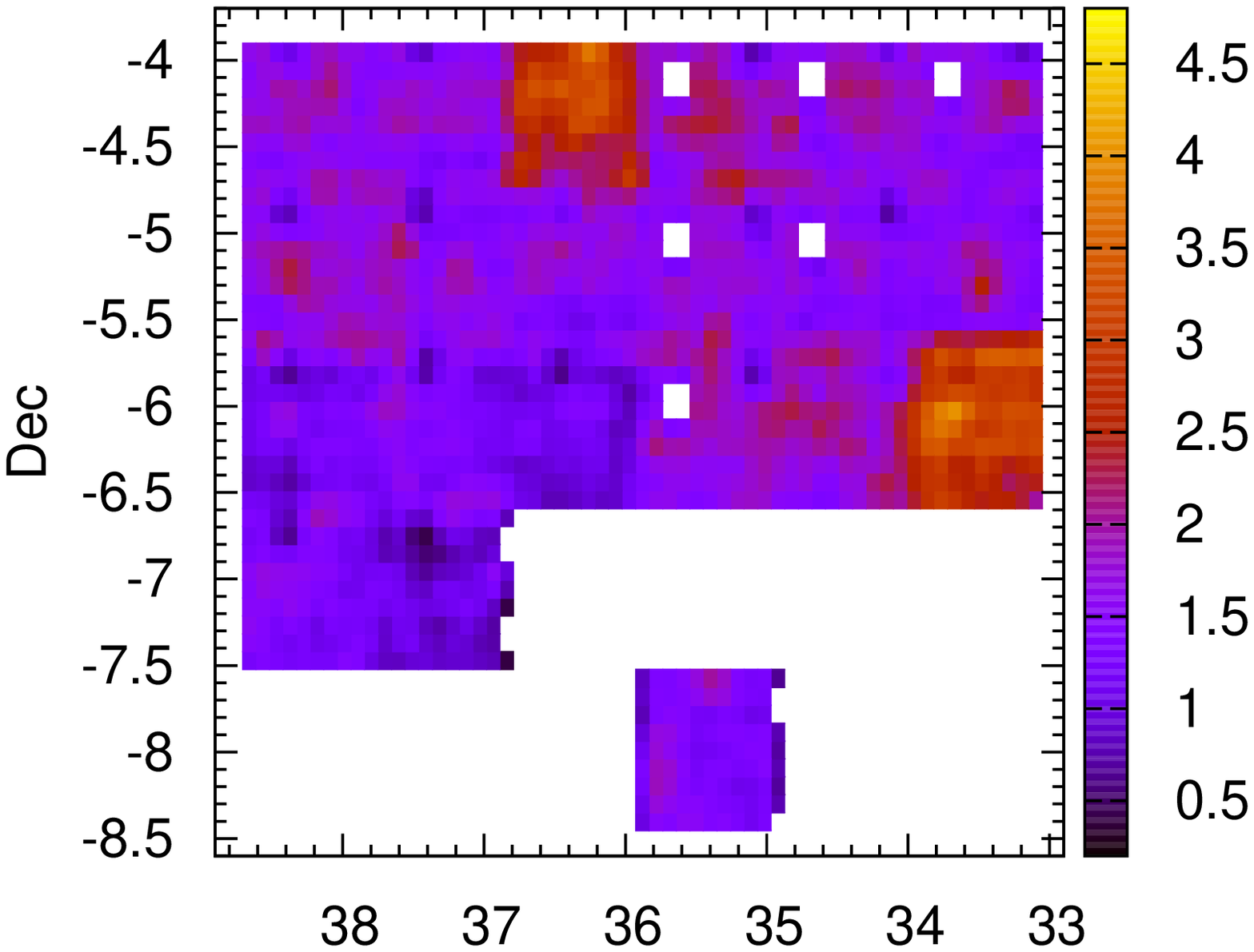}
  \includegraphics[angle=-90,width=0.32\hsize]{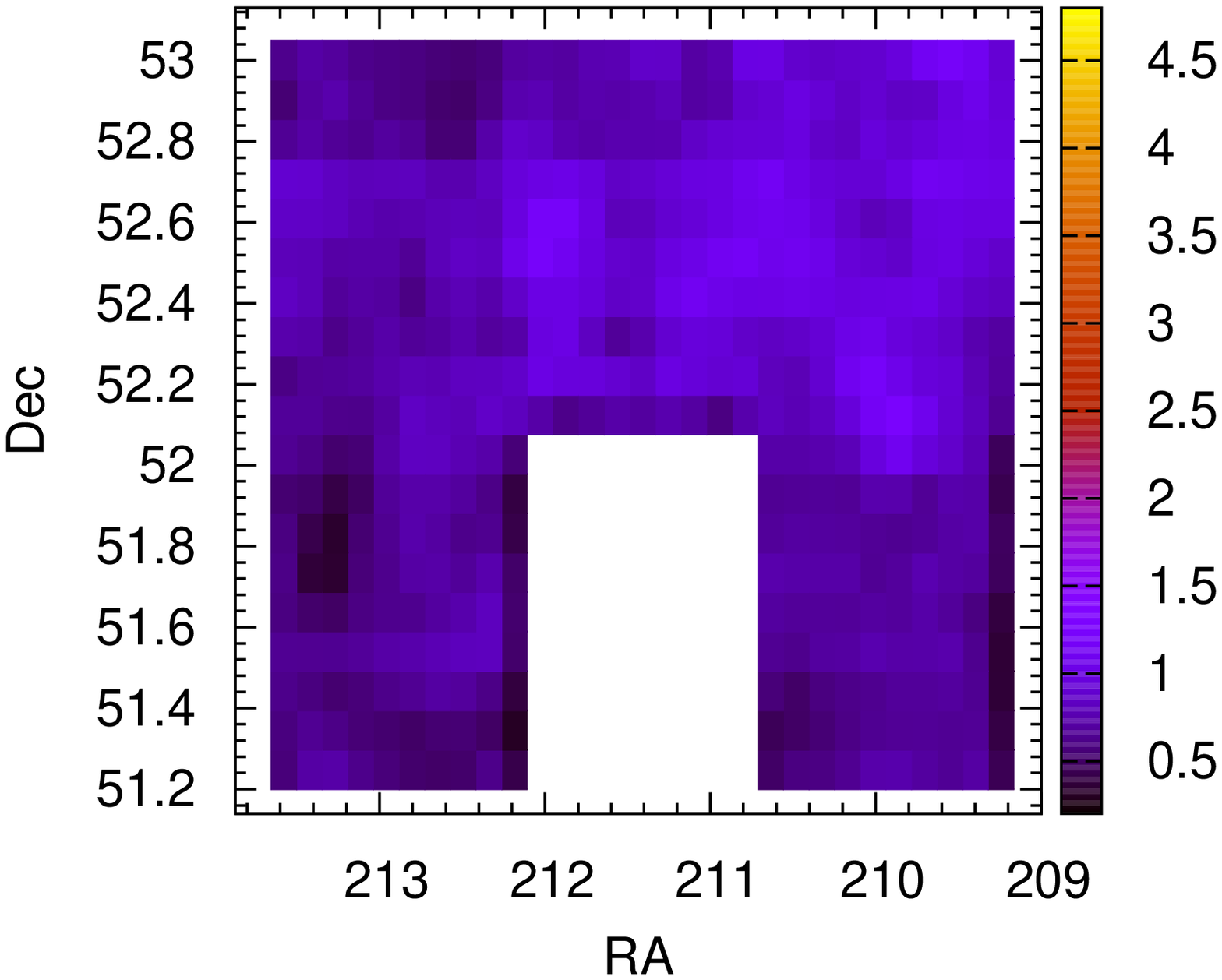}
  \includegraphics[angle=-90,width=0.32\hsize]{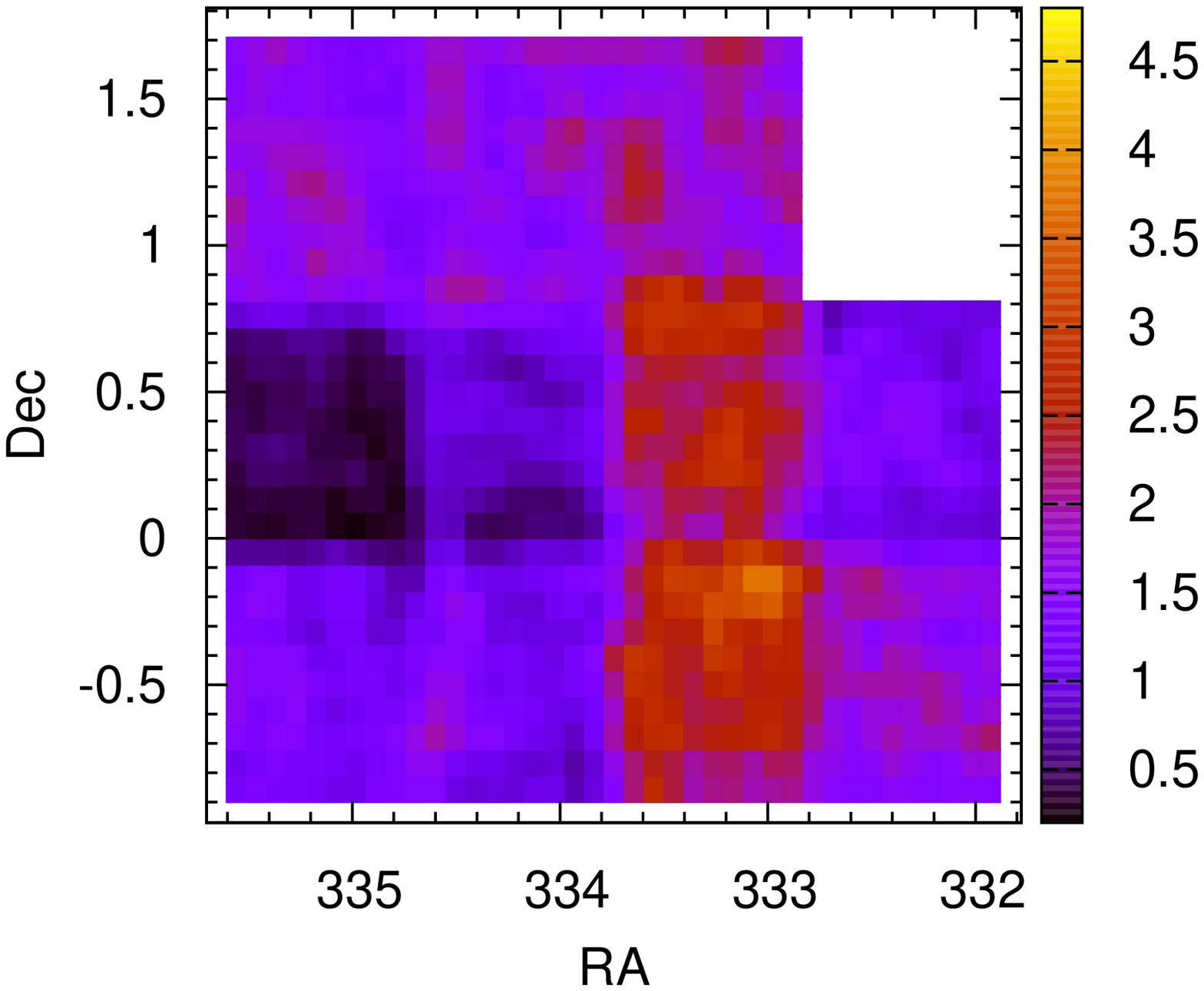}

  \caption{Maps representing the number density per $arcmin^2$ of all
    objects identified by the arcfinder before applying geometrical
    and colour selection criteria, which will be applied in subsequent
    steps to identify the arc candidates. Since the more stringent
    constraints are not yet applied, these maps are dominated by
    spurious detections, thus showing their distribution in the
    survey. The fluctuations in number density are expected because of
    the intrinsic inhomogeneous distribution of the sources and the
    large inhomogeneities of the image depth across the survey. We
    cope with this by measuring the noise locally. The three CARS
    fields, W1, W3 and W4, are shown.}
  \label{fig:carsFields}
\end{figure*}

The depth of images (AB magnitude), defined as the $5\sigma$ detection
limit in a $2^{\prime\prime}$ radius aperture, is typically 25.24,
25.30, 24.36, 24.68 and 23.20 for the $u^\star$, $g^\prime$,
$r^\prime$, $i^\prime$, and $z^\prime$ bands, respectively. The
measured seeing for all co-added images is well below 1.0 arcsecond in
all bands, except for the W1p3p3 $u^\star$-band image, having a seeing
of 1.1 arcseconds, which therefore was ignored in this work. The
seeing quality is crucial to detect arcs because of their very small
width. A large seeing would strongly affect their signal-to-noise
ratio and reduce the $L/W$ ratio, which being their most distinctive
signature would make the detection significantly more difficult.

In addition to the publicly available data, we use the $g^\prime$ and
$r^\prime$ bands to produce a weighted co-addition for each field to
obtain a higher signal-to-noise ratio image, over which we run the
arc\-finder segmentation, as will be discussed later. In the
co-addition, we ignored the $u^\star$ and $z^\prime$ bands because of
their smaller intrinsic depth, and the $i^\prime$ band because of
large elliptical galaxies being too prominent.  Arcs tend to be close
to such objects and would likely blend with them, strongly affecting
their segmentation.

\section{Image segmentation}\label{sec:arcfinder}

As a first step of searching for gravitational arcs, we adopted the
arcfinder described in \citet{2007A&A...472..341S} to produce the image
segmentation necessary to detect elongated objects, and obtain
important geometrical information, such as their length, $L$, and
length-to-width ratio, $L/W$. Both are fundamental quantities for a
subsequent object selection. In this work we briefly summarize the
arcfinder algorithm in Appendix~(\ref{app:arcfinder}).

We calibrated the arcfinder parameters on an empirical basis by
selecting cut-outs centred on arcs previously discovered in the CARS
data, and by adjusting parameters so that arcs are recognised as
such. At the same time, we controlled and evaluated the number of
spurious detections to minimize their presence. At this stage, we set
the final segmentation parameters but defined only very relaxed
geometrical constraints, i.e. length and length-to-width ratio, for
the very first selection of suitable sources. This is to obtain a wide
overview of statistics of their geometrical properties to gain the
necessary information for the actual calibration of the arcfinder as
discussed in the next paragraphs. As an additional rejection criterion
we removed circular areas surrounding bright stars, which might create
false detections because of their halos and diffraction spikes. The
radius in pixels, $r=420\,f^{1/3}$, of these areas is related to the
stellar flux, $f$, as derived from the UCAC3 catalogue
\citep{2010AJ....139.2184Z}. The same catalogue is also used to define
their locations in the field of view. Moreover, we set an upper limit
in the surface brightness of candidates, because arcs tend to be
faint. This limit is set to $0.6 ADU/s$, and was empirically chosen
given the surface brightnesses of the arcs used in the calibration.
The most important parameters used for the segmentation are listed in
Appendix~(\ref{app:arcfinder}). These parameters could be refined by
iterating the process of inclusion of new arcs discovered at each
step. This might be strictly necessary if none or only few known arcs
are present in the field under investigation. In our case we have a
sufficiently large sample for a proper and robust calibration. In
fact, with a unique set of parameters, we detected arcs with very
different shapes, length-to-width ratios, curvature and environment,
ranging from cluster to galactic arcs.

\begin{figure*}[!t]
  \centering
  \includegraphics[angle=-90,width=0.47\hsize]{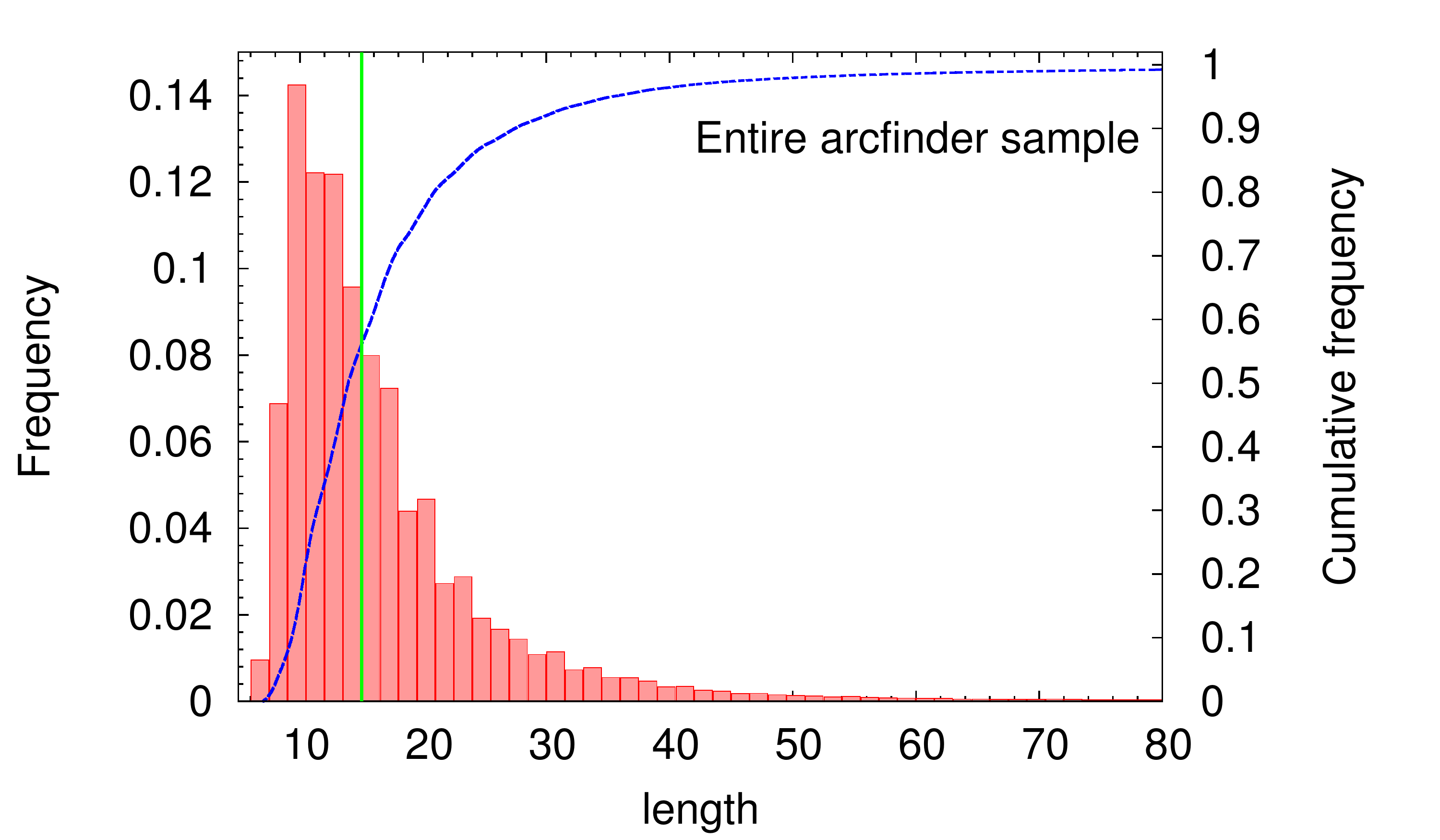}
  \hspace{0.02\hsize}
  \includegraphics[angle=-90,width=0.47\hsize]{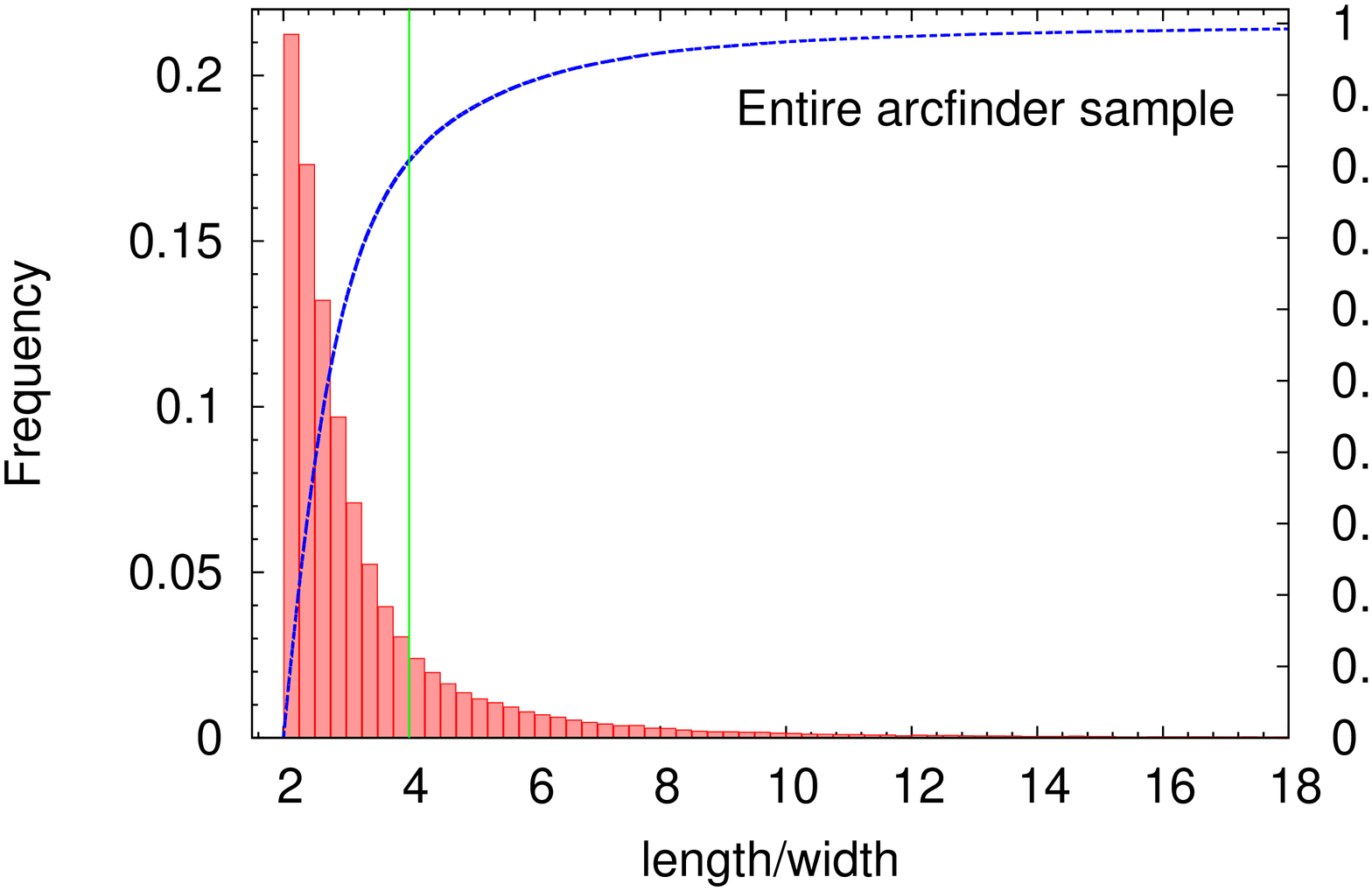}
  \caption{Probability distribution (filled bars) and cumulative
    function (blue line) of the length, expressed in pixels of $0.186$
    arcsec in size, and the length-to-width ratio of all objects
    detected with the arcfinder before the application of the final
    geometrical and colour selection, in the left and right panels,
    respectively. The vertical green lines represent the respective
    lower limits adopted for the final candidates selection to
    maximize the completeness given the limits imposed by the data
    PSF.}
  \label{fig:distri}
\end{figure*}

In Figure~(\ref{fig:arcsegment}) we present a small portion of one of
the cut-outs used for the calibration together with the related
segmentation produced by the arcfinder. Clearly, only very few objects
appear on the right panel of the figure. Detected objects include a
giant arc in the middle, one object facing edge-on, and an elongated
feature resulting from blending of two objects. Blending of the lower
right detection may appear excessive, given the large separation of
the two objects, but we have to keep in mind that the segmentation is
not based on a brightness cut-off criterion as in the case of many
other algorithms. In ours, the segmentation algorithm measures
coherent patterns in local second brightness moments which, for these
two objects, is aligned along a common direction resulting in their
merging as a single detection. We use an aggressive parameter for
blending to avoid splitting of arcs with strong luminosity variations
along their major axis.

With the discussed segmentation and very loose filtering applied so
far to the CARS data, with very relaxed geometrical constraints used
to acquire a wide statistical understanding of the data (the final
selection still has to come), we obtained an initial sample of $
201\,699$ sources, the number density of which across the survey is
shown in Figure~(\ref{fig:carsFields}). The clear fluctuations in
number density follow the intrinsic variations of the object
distribution and, more importantly, depend on the large variations of
the image depth across the survey field.  We account for these
variations during the arcfinding process, where we perform a local
estimate of the noise level to retrieve a sample with uniform
signal-to-noise ratio properties. Note that this initial catalogue is
largely dominated by spurious objects, since the most stringent
constraints are not yet applied. In Figure~(\ref{fig:distri}) we plot
the probability distribution and cumulative function of the main
geometrical properties, i.e. length and length-to-width ratio, of the
entire sample of sources obtained with the arcfinder.

In this step, we aim at completeness by setting only weak constraints
on the detection shapes we expect from strong lensing in combination
with the PSF convolution, hence we choose a minimal length-to-width
ratio of $L/W>4$ instead of bolder values, e.g. $7$ or $10$, typical
for well resolved strong lensing features. Lower values cannot be
adopted, as they are typical for most of the sources in the field
(e.g. galaxies, amongst others). The lack of an additional length
constraint would allow for objects smaller than the image PSF. Because
of this reason, we set a minimum length of $L>15$ pixels (2.9
arcsec). These constraints are shown as green vertical lines in
Figure~(\ref{fig:distri}).  With these selection parameters we reduced
the sample size to $36\,026$ objects, i.e. an average of about $970$
detections per square degree. This is the final sample as provided by
the arcfinder when applied onto a single band. Even if the number of
candidates seems large (and in fact it still contains a large number
of false detections), we have to keep in mind that we started with
$7\: 10^6$ objects detected with SExtractor within the survey
\citep{1996A&AS..117..393B} and that we search for objects down to the
noise limit. This shows how, with the arc segmentation alone, we
already have an efficient and powerful filter, even if not sufficient
to obtain a reliable automatic procedure. In the next section we show
how expanding the method onto three bands improves the situation.

\section{Colour properties of the sample}\label{sec:colors}

Once we obtain the source segmentation, curvature, length and $L/W$
ratio, we get a sample of objects with the right geometrical
properties, i.e. faint, thin and elongated sources. This is however
still not enough for a clear distinction between gravitational arcs
and other astrophysical sources or noise fluctuations. In this work we
address this issue with photometry, in particular of colours. We show,
in fact, that the population of sources with the largest probability
to experience strong lensing (thus resulting in gravitational arcs) is
dominated by a fairly uniform population of galaxies at redshift
$z\approx 1$. These objects are mainly small galaxies with large star
forming regions and therefore specific colour properties
\citep{willmer06}.

\subsection{Theoretical motivation} \label{sec:theory}

\begin{figure*}[!t]
  \centering
  \includegraphics[angle=-90,width=0.48\hsize]{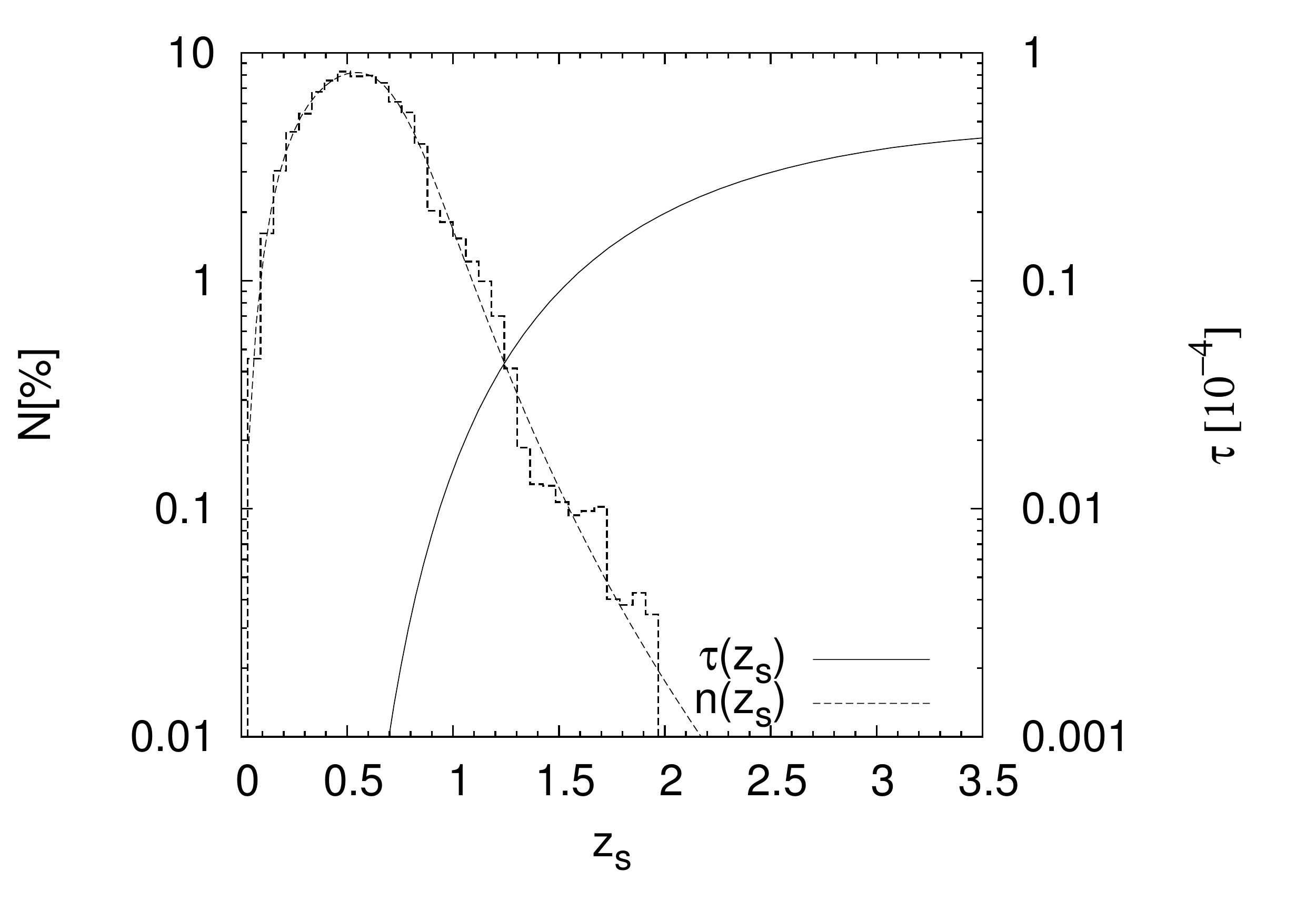}
  \includegraphics[angle=-90,width=0.48\hsize]{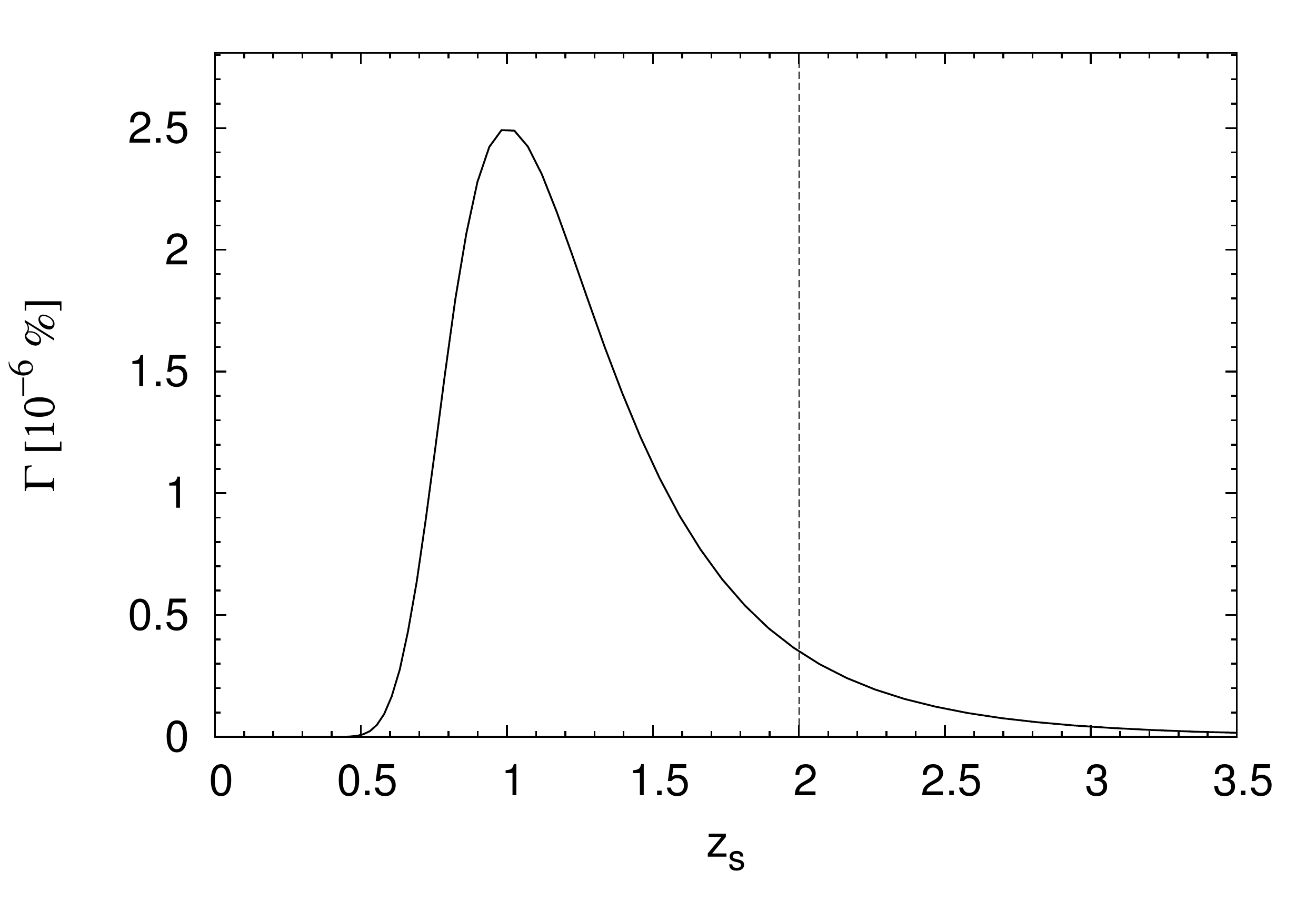}
  \caption{Left panel: Lensing optical depth, $\tau_{\rm d}$, based on
    NFW haloes following the \cite{2002MNRAS.329...61S} differential
    mass function, together with the actual data and best fit of the
    CARS source redshift distribution, $n(z_{\rm s})$. Right panel:
    redshift distribution of the expected number density of arcs,
    $\Gamma_{\rm d}$, based on the best fit of the sources' redshift
    distribution. Note that the best fit of the source distribution
    used in this model overestimates the actual counts which drop for
    $z>1.3$ and vanish at $z\approx 2$, thus suppressing the high
    redshift tail of $\Gamma_{\rm d}$. The vertical line marks the
    redshift limit of the observable sources.}
  \label{fig:proability}
\end{figure*}

The probability of observing sources with strong lensing features is
given by their number-density distribution dependent on redshift,
\begin{equation}
  n(z)=\frac{z^a}{z^b+c\,z^2+d} \;,
\end{equation}
where $a=1.79$, $b=0.17$, $c=8.62$, $d=0.03$, which is the redshift
distribution proposed by \cite{2007MNRAS.381..702B} with an additional
term in the denominator, and the lensing cross section of an
intervening lens,
\begin{equation}
  \sigma_{\rm d}(z_s)=\eta^2\int_{B_{\rm l}}
  \frac{\d^2x}{|\mu(x)|} \;,
\end{equation}
defined as the area on the source plane, where sources are imaged as
arcs with $L/W$ ratio larger than a given minimum, $d$. Here $\mu$ is
the lensing magnification as introduced in
Equation~(\ref{eq:magnification}), $B_{\rm l}$ represents the area in
the lens plane for which the condition $L/W>d$ is met,
$x=\theta/\theta_0$ are dimensionless coordinates scaled by $\theta_0$
and $\eta_0=(D_{\rm s}/D_{\rm l})\theta_0$ in the lens and source
plane, respectively. For simplicity we assume point sources and
spherically symmetric lenses modelled as NFW halos
\citep{1997ApJ...490..493N} with scale radii $r_{\rm s}$. The redshift
distribution of the number density of expected arcs, $\Gamma_{\rm
  d}(z_{\rm s})=n(z_{\rm s})\;\tau(z_{\rm s})$, is obtained by
multiplying the sources number density, $n(z)$, with the sum of the
cross section of all lenses between the observer and the sources
divided by the area of the source plane, i.e. the lensing optical
depth,
\begin{equation}
  \tau_{\rm d}(z_s) = \frac{1}{4\pi D_s^2} \int^{z_{\rm s}}_0 \int^\infty_0
  N(m,z)\, \sigma_{\rm d} (m,z,z_s)\, \d z \d m \;.
\end{equation}
Here, $N(m,z)\d z$ is the total number of haloes with mass $m$
enclosed in the cosmic volume within redshifts $z$ and $z+dz$ as
defined by the \cite{2002MNRAS.329...61S} differential mass function.
In the left and right panels of Figure~(\ref{fig:proability}) we show
the sources distribution of the CARS galaxy sample together with the
lensing optical depth, $\tau_{\rm d}$, and the resulting distribution
of the arcs, $\Gamma_{\rm d}$, respectively. Here, we can safely use
the redshift distribution of all sources in the survey, which is
dominated by weakly lensed sources, and fully ignore their flux
enhancement, which would be caused by a strong lens. This is because
while lensing does increase the total flux of a source, it does not
affect its surface brightness, which is what the initial arcfinder
detection is more sensitive to. The peak of the number of strongly
lensed sources at $z\approx 1$, is caused by the steep rise of the
optical depth and the drop of the source number density with
increasing redshift. The high redshift tail of our model overestimates
the actual number of expected arcs because the best fit of the sources
overestimates the actual data, which drop for $z>1.3$ to become zero
for $z>2$. For this reason, we can assume that the largest number of
strongly lensed sources in our data are confined to a relatively
narrow redshift interval around $z\approx 1$.

Many different works based on numerical N-body simulations and
halo-models have been devoted to the evaluation of the lensing optical
depth with extended sources, the lens intrinsic ellipticity and the
presence of substructures within large haloes \citep[see
  e.g.][]{2003MNRAS.340..105M,2005ApJ...635..795L,
  2006A&A...447..419F}. These additional details mostly increase the
efficiency of lenses to produce arcs and only marginally affect its
redshift dependence, which we are interested in. With our simple model
we just focus the attention on the sources rather than on the lenses,
in contrast to most of these studies. It is neither meant to produce a
detailed prediction of the number of observable arcs nor to define the
actual parameters used to perform the object identification. Here, we
highlight the priciples which motivate our selection criteria aiming
at the objects with the highest probability to be strongly lensed,
i.e. those at $z\approx 1$. The actual selection will be performed by
calibrating the method on the colours of known arcs as it will be
detailed in the following section.

\subsection{Observational evidence}

To substantiate our line of argument, we measured photometric
properties of the entire sample for all available bands. Magnitudes
were measured via aperture photometry
\begin{equation}
  m = ZP - 2.5\;\log_{10}( I - S\, A ) \;,
\end{equation}
where $ZP$ is the zero-point magnitude (as given in the CARS fits
headers), $I$ is the total flux in the aperture given in ADUs, $A$ is
the number of pixels within the aperture, and $S= 2.5\;median -
1.5\;mean$ is the background mode \citep{1996A&AS..117..393B}
evaluated on the area enclosed within $4.5$ arcsec from the outer edge
of the aperture. The so-defined background correction accounts for a
possible flux contribution of the lens candidate, which should be
close to the arc it produces. This implicitly assumes that the lens
light profile decreases linearly in the immediate vicinity of the arc,
as justified by its small width.  We remind the reader that the
aperture of each individual object is defined by the segmentation
produced with the arcfinder from the weighted stack of the $g^\prime$
and $r^\prime$ images (see Section~\ref{sec:dataset}), and is kept
fixed for all bands.

\begin{figure*}[!t]
  \centering
  \includegraphics[angle=-90,width=0.45\hsize]{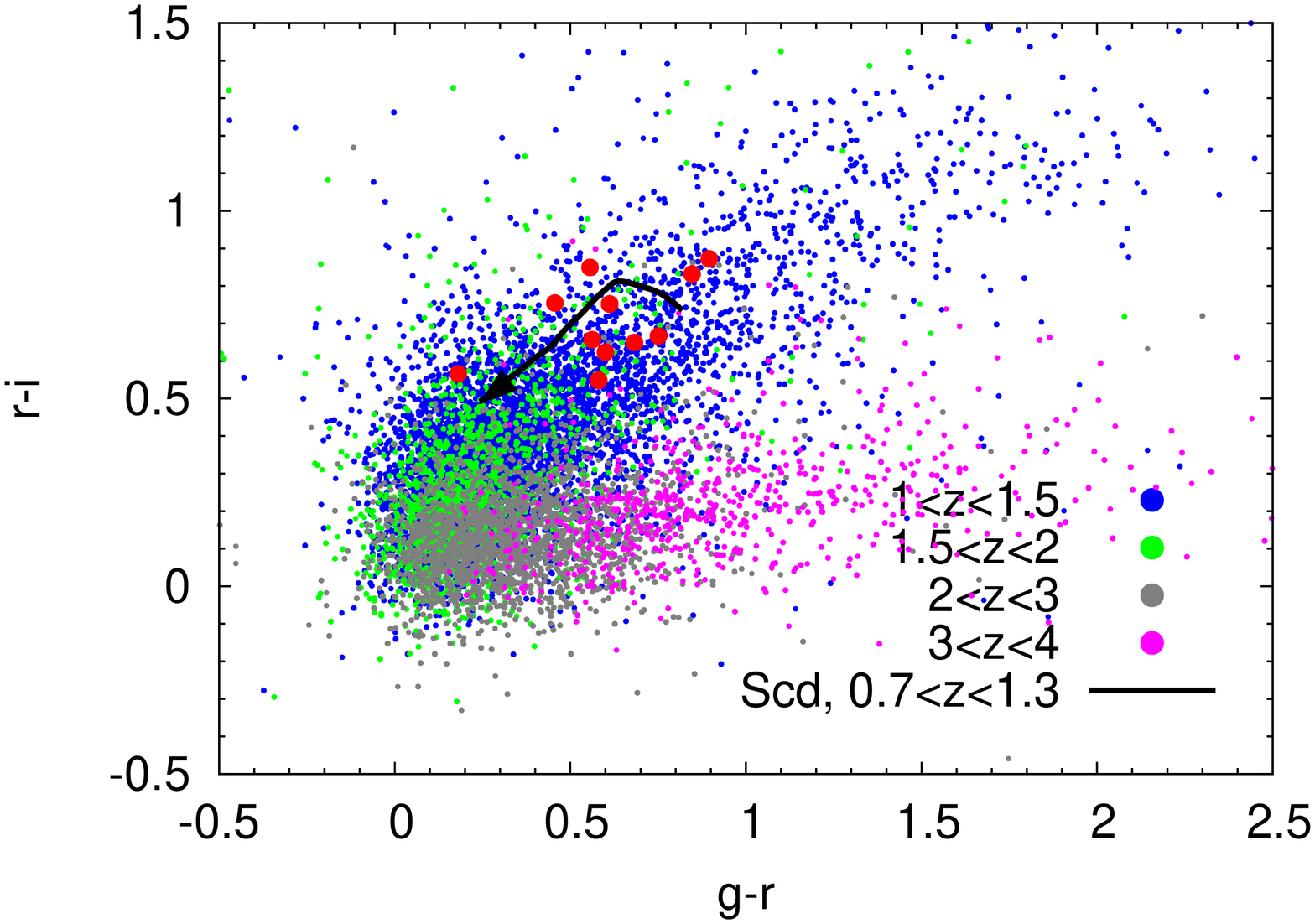}
  \includegraphics[angle=-90,width=0.505\hsize]{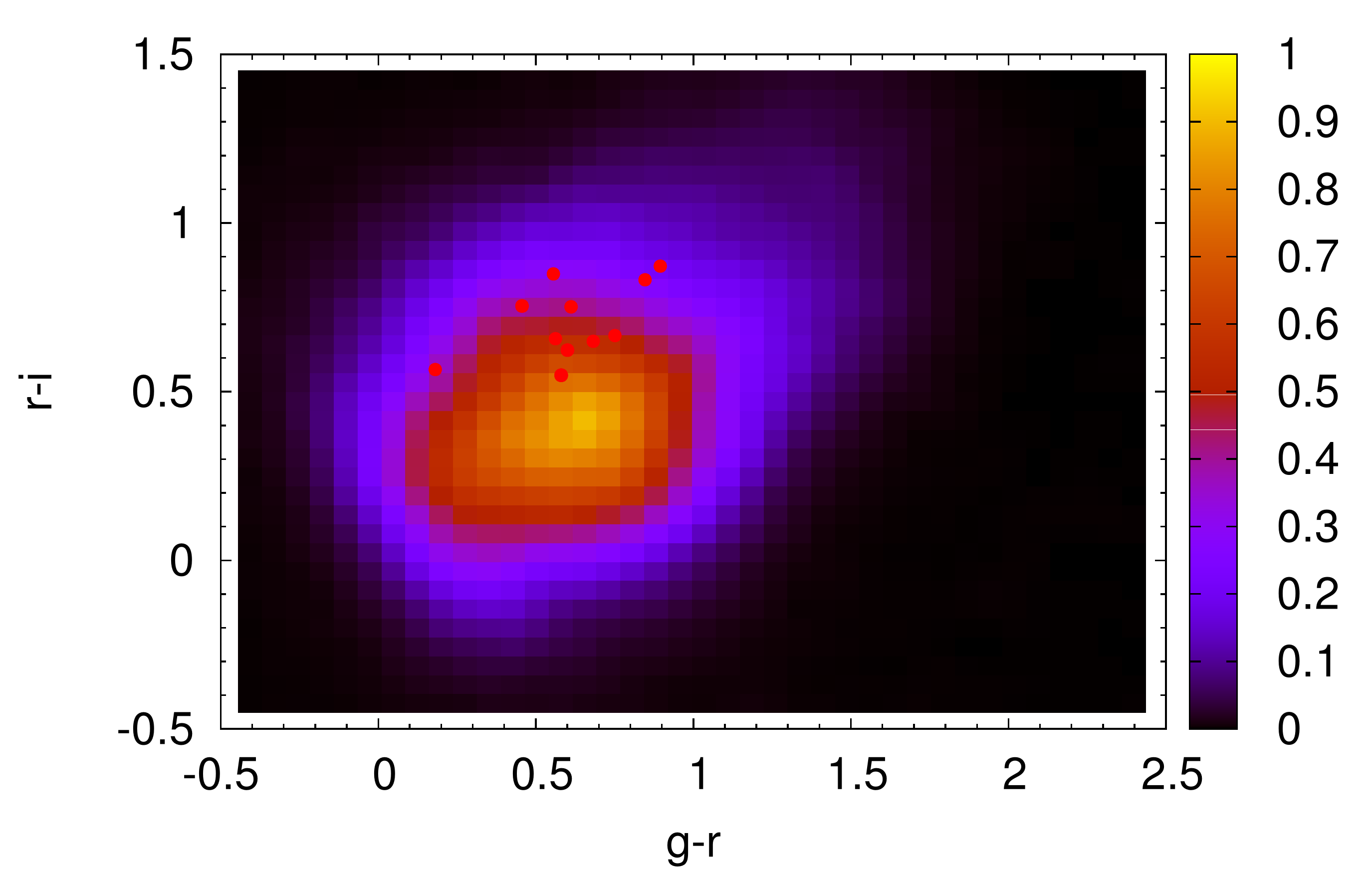}
  \caption{Left panel shows a colour-colour diagram for previously known
    arcs, in red, and a sub-sample of galaxies observed in the COSMOS
    field subdivided in redshift bins. The red points denote the arcs
    previously known in the survey and which are used to calibrate the
    colour selection. The small colour region where arcs are present
    shows how the lensed sources lie at redshift $z\approx 1$ (as
    predicted by our optical depth model) and have well defined colour
    properties (as shown by the black arrow referring to a Scd galaxy,
    for which we derived the photometry out of a synthetic SED, here
    the arrowhead points in the direction of increasing
    redshifts). The right panel shows the same colour-colour diagram,
    but for all objects detected by the arcfinder.}
  \label{fig:gr-ri}
\end{figure*}

In the left panel of Figure~(\ref{fig:gr-ri}) we plotted a
colour-colour diagram ($g-r$,$r-i$) showing, as red circles, the most
evident arcs already known in the literature
\citep{2012ApJ...749...38M} for which we could derive photometry based
on the arcfinder segmentation. The other points refer to a sub-sample
of galaxies, grouped in redshift bins, observed with the SUBARU
telescope in the COSMOS field \citep{2009ApJ...690.1236I}. The colour
redshift dependence of the galaxies is clear, and shows how lensed
sources are indeed associated to galaxies at redshift $z\approx 1$, as
expected. The black arrow represents colours of a Scd galaxy, with
large spiral arms dominated by a population of young stars for
redshifts ranging from $z=0.7$ (start of line) to $z=1.3$ (arrowhead),
as derived from a synthetic spectral energy distribution (SED)
produced by \cite{1980ApJS...43..393C}. This is not used to define the
colour cuts but to show the region where we expect to find galaxies
with bright star-forming regions and therefore well defined colour
properties. In the right panel of Figure~(\ref{fig:gr-ri}) we repeated
the same exercise, but with all sources detected by the arcfinder. It
is visible how the covered by the arcs excludes a very large
fraction of sources (number density is shown in the background),
helping to drastically reduce sample contamination. With these
arguments, supported by our theoretical model and data alike, we now
have a robust basis to help us in distinguishing arcs from other
astrophysical sources. We favour this simple colour based method in
contrast to a full photometric redshift estimate, because the redshift
catalogues available in the literature are not well tailored for the
task. This is because usually adopted segmentation and de-blending are
likely to merge arcs with relatively bright galaxies to which they are
associated, resulting in misleading and largely incomplete results.
The colour selection is sufficient to constrain the redshift range
which has to be investigated.

We can now proceed to define colour selection, where for convenience we
use the flux related quantity
\begin{equation}\label{eq:flux}
  f_x=A^{-1}\,10^{\,0.4\,(K-m_x)} \;,
\end{equation}
to express the difference in flux over area for neighbouring bands in
the fashion of usual colour-colour plots, instead of the typical
definition based on magnitudes. Here, $m_x$ is the object magnitude in
$x$-band, where $x$ stands for
$x=[u^\star,g^\prime,r^\prime,i^\prime,z^\prime]$, and $K=26.42$ is
the magnitude average of all sources in the $r^\prime$-band resulting
in a multiplicative factor introduced for convenience. 

\begin{figure*}[!t]
  \centering
  \includegraphics[angle=-90,width=0.98\hsize]{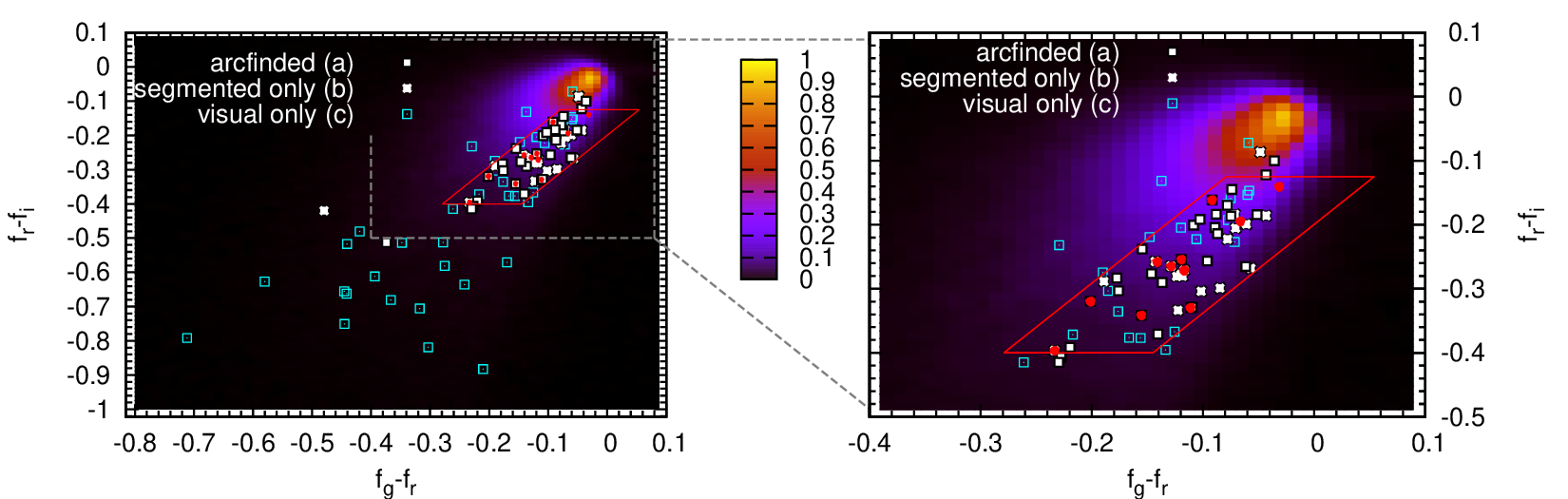}
  \caption{Flux related ``colour-colour'' space for all arcfinder
    detections (background intensity), the previously known arcs
    present in the field for which we derived their photometry (red
    points), and the region, within the red lines, used to select the
    final arcs sample. The white squares (a) are the candidates
    detected by the arcfinder, the white asterisks (b) are those
    segmented by the arcfinder but rejected by its geometrical filters
    and the cyan void squares (c) are those found purely by visual
    inspection. The latter are the most difficult sources and are
    scattered in a very wide area, far from most of the other sources,
    because of their uncertain photometry.}
  \label{fig:gr-ri_flux}
\end{figure*}

In this ``colour-colour'' space, based on Equation~(\ref{eq:flux}),
the arcs previously known in the survey, and plotted in
Figures~(\ref{fig:gr-ri}) and (\ref{fig:gr-ri_flux}) as red circles,
appear nicely aligned along a relatively narrow region in
colour-colour space. These objects were used to define the region,
marked with red lines in Figure~(\ref{fig:gr-ri_flux}), that we used
to select other arc candidates. In the same figure we also plot the
objects we identified in the survey as arcs, where details will be
given in Section~(\ref{sec:visual}). For the moment, it shall suffice
to say that all objects marked in these plots were not selected
according to their colour properties, making them an independent check
sample.  These detections as well predominantly populate the same
colour-colour space we used to define the colour selection, lending
further support to the validity of our approach.

It is difficult to extract detailed statistics from the small sample
of known arcs, hence we decided not to use a more sophisticated colour
selection criterion. The sample of detections after the colour
selection reduces to $5\,597$ candidates, i.e. approximately 150 per
square degree, versus 970 per square degree in the catalogue obtained
with the single-band arcfinder, with comparatively less
contamination. This colour selection clearly restricts the detection
of arcs to a certain type of lensed sources. Nevertheless, we do not
consider this as a limitation, because (1) we expect the population of
lensed sources selected in this way to be the most numerous, ensuring
that only a small fraction of arcs is lost, as will be shown later in
the text, (2) sources and lenses are completely uncorrelated, because
of their large relative separation, avoiding any bias which may be due
to physical correlations, and (3) gravitational lensing is a
completely achromatic phenomenon independent of the colour of the
sources. The giant arcs count, used to infer cosmological information,
depends on this colour selection, but this can be easily accounted for
just by applying the same colour selection to the expected number
density of background sources, which is a trivial task.

\section{Visual validation and catalogue}\label{sec:visual}

The candidates produced with the discussed procedure are finally
verified by visually inspecting their colour composite images obtained
using the $g^\prime$, $r^\prime$ and $i^\prime$ bands. We associated a
rank to each detection based on three levels:
\begin{itemize}
\item {\it 1=unlikely}: arc structures which seem physically
  associated to an astrophysical object, or which are relatively far
  from a possible lens,
\item {\it 2=unclear}: arc structures with a very promising shape, but
  still of dubious origin,
\item {\it 3=very likely}: a clear curved arc structure around a
  possible lens causing its origin, i.e. large ellipticals or
  significant concentrations of galaxies.
\end{itemize}
A rank $0$ is also used to label artefacts, evident spiral galaxies,
or clear noise fluctuations. This simple ranking yields stable results
with respecto to different observers and even the same observer but at
different times.  In addition we also inspected the full survey to
evaluate the completeness and purity of the sample against a full
visual investigation. This very important topic will be covered in
Section~(\ref{sec:complete}).

The final catalogue listing all arcs with rank $r\geq 2$ is shown in
Tables~(\ref{tab:catalog}, \ref{tab:catalog1} and \ref{tab:catalog2})
with their positions, geometrical properties, photometry, redshifts
and ranks. The first two tables contain the 56 detections satisfying
the colour criterion, while the 34 detections in the third table, most
of them found only by visual inspection, do not. The candidates
detected by the arcfinder, i.e satisfying the tight geometrical
constraints, are labelled with `a'. Only five of them do not meet the
colour criterion. Three of those (arc58a, arc59a, arc60a) are located
in the vicinity of very bright stars whose halos might have spoiled
their photometry, and one (arc61a) belongs to a complex object of
unclear nature, possibly showing multiple arcs. The objects segmented
by the arcfinder but discarded by it because they do not fulfil the
geometrical constraints are labelled with `b' and all but three of
them satisfy the colour constraints. Finally, the candidates detected
via eye-ball checking the whole survey and missed entirely by the
arcfinder are labelled with `c'. The majority of the visually
identified objects fall into Table~(\ref{tab:catalog2}) of detections
rejected by the colour selection because of their low surface
brightness and heavy blending, which makes them generally the most
difficult candidates to find.  The complete sample consists of 90
objects, 73 of which are newly proposed arc candidates. Postage stamp
cut-outs of their images are available
online\footnote{\url{http://www.ita.uni-heidelberg.de/~maturi/Public/arcs}}.

We obtained more candidates with respect to those proposed by
\cite{2012ApJ...749...38M} and \cite{2007A&A...461..813C} not only
because of the different arcfinder adopted but also because of the
different selection criteria and because they restricted their sample
with respect to the arc-lens separation to $r>14$ arcsec. Note that we
dropped two of their candidates, i.e. SA98 and SL2SJ021932-053135,
which in our opinion are not lensing features but structures
physically associated to the galaxy believed to be the lens in the
first case, and a spiral arm of a nearby galaxy in the latter. The
arcfinder used in this work clearly detected SA38, but this was
rejected because of its high luminosity exceeding the maximum we
allowed.

\begin{sidewaystable*}
  \vspace{18cm}  
  \centering
  \caption {Final gravitational arc candidates satisfying the colour
    criteria. Detections with $RANK\le1$, i.e. clear artefacts or
    features possibly related to spiral galaxies or tidal streams, are
    not listed. The objects labelled with `a' were detected by the
    arcfinder Those labelled with `b' were segmented by the arcfinder
    but rejected by our tight geometrical constraints. Those labelled
    with `c' were detected only via eye-ball checking. The table
    lists: the arc identification number `ID', the CARS field `FIELD',
    the Right Ascension and Declination, `RA' and `DEC', the arc
    length `LEN' in arcsec, the length-to-width ratio `L/W', the
    curvature `CUR' in $\rm arcsec^{-1}$, the area `A' in $\rm
    arcsec^2$, the $u^\star$,$g^\prime$,$r^\prime$,$i^\prime$
    magnitudes, the lens redshift `$z_{\rm lens}$', the arc candidate
    redshift `$z_{\rm arc}$', the detection ranking `RANK', and the
    identification number in \cite{2012ApJ...749...38M} or
    \cite{2007A&A...461..813C} if present. The arc redshift is
    enclosed within brackets to stress its low reliability.}
  \label{tab:catalog}
  \begin{tabular}{llrrrrrrrrrrrrr}
    ID & FIELD & RA & DEC & LEN & L/W & CUR & A & m$_{u\prime}$ & m$_{g\prime}$ & m$_{r\prime}$ & m$_{i\prime}$ & z$_{\rm lens}$ & RANK & CAT\\
    \hline
arc1$_a$   & W1m1p1 & 02:13:17.2 & -06:25:58 & 2.91 & 4.66 & 0.21 & 2.0 & $25.6^{\pm 0.6}$ & $24.3^{\pm 0.2}$ & $23.7^{\pm 0.5}$ & $23.1^{\pm 0.2}$ & $0.39_{-0.09}^{+0.09}$ & 2 & - \\
arc2$_a$   & W1m1p2 & 02:14:08.0 & -05:35:27 & 7.92 & 8.65 & 0.14 & 8.1 & $23.0^{\pm 0.2}$ & $22.3^{\pm 0.1}$ & $21.6^{\pm 0.1}$ & $21.0^{\pm 0.1}$ & $0.49_{-0.21}^{+0.22}$ & 3 & SA22 \\
arc3$_a$   & W1m1p1 & 02:14:40.5 & -06:31:16 & 4.56 & 5.06 & 0.18 & 4.0 & $24.7^{\pm 0.3}$ & $23.8^{\pm 0.1}$ & $23.1^{\pm 0.2}$ & $22.3^{\pm 0.1}$ & $0.34_{-0.09}^{+0.09}$ & 2 & - \\
arc4$_a$   & W1m1p1 & 02:16:00.2 & -05:57:04 & 4.08 & 4.82 & 0.18 & 3.6 & $25.1^{\pm 0.4}$ & $24.0^{\pm 0.2}$ & $23.1^{\pm 0.3}$ & $22.4^{\pm 0.1}$ & $0.36_{-0.09}^{+0.09}$ & 2 & - \\
arc5$_a$   & W1m0p3 & 02:17:13.2 & -04:01:13 & 4.28 & 5.92 & 0.73 & 3.5 & $24.7^{\pm 0.4}$ & $24.2^{\pm 0.2}$ & $23.5^{\pm 0.3}$ & $22.6^{\pm 0.1}$ & $0.68_{-0.11}^{+0.11}$ & 2 & - \\
arc6$_a$   & W1m0p2 & 02:18:05.0 & -05:19:49 & 5.53 & 4.23 & 0.28 & 6.5 & $23.2^{\pm 0.2}$ & $22.8^{\pm 0.1}$ & $22.4^{\pm 0.1}$ & $21.5^{\pm 0.1}$ & $0.77_{-0.12}^{+0.12}$ & 2 & - \\
arc7$_a$   & W1m0p1 & 02:19:03.9 & -05:51:25 & 4.03 & 5.15 & 0.00 & 3.7 & $25.1^{\pm 0.6}$ & $24.2^{\pm 0.2}$ & $23.3^{\pm 0.2}$ & $22.3^{\pm 0.1}$ & $0.44_{-0.09}^{+0.09}$ & 2 & - \\
arc8$_a$   & W1m0p3 & 02:19:09.6 & -04:01:43 & 5.66 & 4.90 & 0.12 & 5.9 & $24.1^{\pm 0.3}$ & $23.3^{\pm 0.1}$ & $22.5^{\pm 0.1}$ & $21.7^{\pm 0.1}$ & $0.33_{-0.09}^{+0.09}$ & 3 & SA35 \\
arc9$_a$   & W1m0p3 & 02:19:37.5 & -04:16:18 & 3.59 & 4.57 & 0.23 & 3.0 & $24.4^{\pm 0.3}$ & $23.9^{\pm 0.1}$ & $23.5^{\pm 0.3}$ & $22.7^{\pm 0.1}$ & $0.49_{-0.21}^{+0.22}$ & 2 & - \\
arc10$_a$   & W1m0p2 & 02:19:56.4 & -05:27:56 & 3.73 & 4.83 & 0.21 & 2.9 & $23.8^{\pm 0.2}$ & $23.0^{\pm 0.1}$ & $22.4^{\pm 0.1}$ & $21.8^{\pm 0.1}$ & $0.32_{-0.09}^{+0.09}$ & 3 & SA36 \\
arc11$_a$   & W1p1p1 & 02:20:44.4 & -05:37:46 & 4.34 & 4.76 & 0.07 & 4.1 & $24.4^{\pm 0.3}$ & $23.4^{\pm 0.1}$ & $22.9^{\pm 0.2}$ & $22.1^{\pm 0.1}$ & $0.72_{-0.11}^{+0.11}$ & 2 & - \\
arc12$_a$   & W1p1p3 & 02:21:34.5 & -04:24:38 & 7.44 & 12.03 & 0.32 & 6.0 & $23.5^{\pm 0.2}$ & $22.6^{\pm 0.1}$ & $22.0^{\pm 0.1}$ & $21.3^{\pm 0.1}$ & $0.70_{-0.11}^{+0.11}$ & 3 & - \\
arc13$_a$   & W1p2p2 & 02:25:32.0 & -04:50:59 & 5.49 & 8.20 & 0.37 & 4.4 & $23.2^{\pm 0.2}$ & $22.6^{\pm 0.1}$ & $21.9^{\pm 0.1}$ & $21.3^{\pm 0.1}$ & $0.62_{-0.11}^{+0.11}$ & 2 & SL2SJ022532-045100 \\
arc14$_a$   & W1p3p1 & 02:31:06.7 & -05:55:02 & 3.52 & 4.50 & 0.51 & 2.8 & $25.1^{\pm 0.5}$ & $24.1^{\pm 0.2}$ & $23.4^{\pm 0.4}$ & $22.8^{\pm 0.2}$ & $0.53_{-0.10}^{+0.10}$ & 3 & SA57 \\
arc15$_a$   & W3m3m2 & 14:00:21.2 & +52:15:47 & 7.21 & 10.00 & 0.40 & 6.2 & $23.1^{\pm 0.2}$ & $22.3^{\pm 0.1}$ & $21.8^{\pm 0.1}$ & $21.1^{\pm 0.1}$ & $0.76_{-0.12}^{+0.12}$ & 2 & - \\
arc16$_a$   & W3m3m2 & 14:01:44.6 & +53:02:09 & 3.67 & 4.20 & 0.13 & 3.5 & $24.0^{\pm 0.2}$ & $23.4^{\pm 0.1}$ & $22.8^{\pm 0.2}$ & $22.0^{\pm 0.1}$ & $0.47_{-0.10}^{+0.10}$ & 3 & SA91 \\
arc17$_a$   & W4m1m2 & 22:09:18.3 & +00:50:03 & 3.17 & 4.16 & 0.45 & 2.4 & $25.3^{\pm 0.5}$ & $23.7^{\pm 0.2}$ & $23.0^{\pm 0.3}$ & $22.4^{\pm 0.1}$ & $0.49_{-0.21}^{+0.22}$ & 2 & - \\
arc18$_a$   & W4m0m0 & 22:12:22.0 & +01:36:46 & 4.08 & 4.61 & 0.37 & 4.4 & $24.5^{\pm 0.3}$ & $23.7^{\pm 0.1}$ & $23.1^{\pm 0.2}$ & $22.4^{\pm 0.1}$ & $0.72_{-0.11}^{+0.11}$ & 3 & - \\
arc19$_a$   & W4m0m1 & 22:12:55.3 & +00:17:07 & 6.37 & 6.34 & 0.24 & 6.7 & $24.0^{\pm 0.2}$ & $23.0^{\pm 0.1}$ & $22.2^{\pm 0.1}$ & $21.4^{\pm 0.1}$ & $0.58_{-0.10}^{+0.10}$ & 2 & - \\
arc20$_a$   & W4m0m2 & 22:13:06.9 & +00:18:30 & 5.34 & 4.26 & 0.20 & 5.9 & $23.8^{\pm 0.2}$ & $22.9^{\pm 0.1}$ & $22.3^{\pm 0.1}$ & $21.7^{\pm 0.1}$ & $0.49_{-0.21}^{+0.22}$ & 3 & - \\
arc21$_a$   & W4m0m0 & 22:13:14.2 & +00:50:09 & 7.80 & 6.08 & 0.15 & 8.2 & $24.6^{\pm 0.3}$ & $23.4^{\pm 0.1}$ & $22.5^{\pm 0.1}$ & $21.6^{\pm 0.1}$ & $0.45_{-0.10}^{+0.10}$ & 2 & - \\
arc22$_a$   & W4m0m0 & 22:14:18.9 & +01:10:35 & 6.14 & 8.76 & 0.60 & 4.9 & $24.4^{\pm 0.3}$ & $23.3^{\pm 0.1}$ & $22.4^{\pm 0.2}$ & $21.6^{\pm 0.1}$ & $0.56_{-0.10}^{+0.10}$ & 3 & SA125 \\
arc23$_a$   & W4p1m0 & 22:15:13.4 & +01:02:41 & 6.85 & 6.62 & 0.07 & 7.3 & $23.8^{\pm 0.3}$ & $22.7^{\pm 0.1}$ & $22.1^{\pm 0.1}$ & $21.2^{\pm 0.1}$ & $0.69_{-0.11}^{+0.11}$ & 2 & - \\
arc24$_a$   & W4p1m2 & 22:16:58.7 & +00:01:53 & 3.99 & 5.65 & 0.00 & 2.9 & $24.2^{\pm 0.4}$ & $23.4^{\pm 0.2}$ & $22.7^{\pm 0.2}$ & $22.1^{\pm 0.1}$ & $0.67_{-0.11}^{+0.11}$ & 2 & - \\
arc25$_b$   & W1m1p3 & 02:14:11.1 & -04:05:03 & 1.68 & 3.24 & 0.33 & 1.2 & $24.5^{\pm 0.4}$ & $24.2^{\pm 0.2}$ & $23.9^{\pm 0.3}$ & $23.1^{\pm 0.2}$ & $0.77_{-0.12}^{+0.12}$ & 3 & SA23 \\
arc26$_b$   & W1m1p1 & 02:14:26.4 & -05:39:39 & 3.73 & 3.56 & 0.19 & 3.8 & $24.6^{\pm 0.3}$ & $23.7^{\pm 0.1}$ & $22.9^{\pm 0.2}$ & $22.0^{\pm 0.1}$ & $0.55_{-0.10}^{+0.10}$ & 2 & - \\
arc27$_b$   & W1m0p2 & 02:16:06.4 & -04:49:35 & 4.19 & 2.53 & 0.52 & 5.7 & $23.8^{\pm 0.2}$ & $23.4^{\pm 0.1}$ & $22.9^{\pm 0.1}$ & $22.0^{\pm 0.1}$ & $0.30_{-0.09}^{+0.14}$ & 3 & - \\
arc28$_b$   & W1m0p2 & 02:17:37.4 & -05:13:30 & 2.72 & 2.76 & 0.21 & 2.8 & $24.8^{\pm 0.4}$ & $24.1^{\pm 0.2}$ & $23.7^{\pm 0.4}$ & $22.8^{\pm 0.2}$ & $0.82_{-0.12}^{+0.12}$ & 3 & SL2SJ021737-051329 \\
arc29$_b$   & W1m0p2 & 02:18:07.4 & -05:15:36 & 2.72 & 3.61 & 0.07 & 2.1 & $25.5^{\pm 0.7}$ & $24.0^{\pm 0.1}$ & $23.4^{\pm 0.2}$ & $22.6^{\pm 0.1}$ & $0.60_{-0.10}^{+0.11}$ & 3 & SA33 \\
arc30$_b$   & W1m0p1 & 02:19:28.3 & -05:44:59 & 2.00 & 2.33 & 0.45 & 1.7 & $25.2^{\pm 0.6}$ & $24.3^{\pm 0.3}$ & $23.9^{\pm 0.4}$ & $23.2^{\pm 0.2}$ & $0.26_{-0.08}^{+0.08}$ & 3 & - \\
    \hline
  \end{tabular}
\end{sidewaystable*}

\begin{sidewaystable*}
  \vspace{18cm}  
  \centering
  \caption {Continued from Table~(\ref{tab:catalog}).}
  \label{tab:catalog1}
  \begin{tabular}{llrrrrrrrrrrrrr}
    ID & FIELD & RA & DEC & LEN & L/W & CUR & A & m$_{u\prime}$ & m$_{g\prime}$ & m$_{r\prime}$ & m$_{i\prime}$ & z$_{\rm lens}$ & RANK & CAT\\
    \hline
arc31$_b$   & W1p1p1 & 02:21:14.9 & -05:42:42 & 1.92 & 2.05 & 0.27 & 2.1 & $24.7^{\pm 0.4}$ & $23.6^{\pm 0.1}$ & $22.9^{\pm 0.3}$ & $22.3^{\pm 0.1}$ & $0.28_{-0.08}^{+0.09}$ & 2 & - \\
arc32$_b$   & W1p1p1 & 02:23:15.3 & -06:29:03 & 2.33 & 3.39 & 0.26 & 2.0 & $24.9^{\pm 0.4}$ & $23.6^{\pm 0.1}$ & $22.9^{\pm 0.1}$ & $22.2^{\pm 0.1}$ & $0.57_{-0.10}^{+0.10}$ & 3 & SA40 \\
arc33$_b$   & W1p2p2 & 02:24:27.3 & -04:55:43 & 1.34 & 2.01 & 0.21 & 1.0 & $25.2^{\pm 0.6}$ & $24.4^{\pm 0.2}$ & $23.9^{\pm 0.5}$ & $23.3^{\pm 0.2}$ & $0.56_{-0.10}^{+0.10}$ & 3 & - \\
arc34$_b$   & W1p4p3 & 02:33:07.2 & -04:38:38 & 2.23 & 3.98 & 0.11 & 1.5 & $25.5^{\pm 0.7}$ & $24.1^{\pm 0.2}$ & $23.6^{\pm 0.3}$ & $22.9^{\pm 0.2}$ & $0.66_{-0.11}^{+0.11}$ & 3 & SA59 \\
arc35$_b$   & W3m3m2 & 13:57:02.4 & +52:30:39 & 2.43 & 4.71 & 0.17 & 1.6 & $25.7^{\pm 0.8}$ & $24.7^{\pm 0.4}$ & $23.8^{\pm 0.6}$ & $23.0^{\pm 0.3}$ & $0.38_{-0.09}^{+0.09}$ & 3 & - \\
arc36$_b$   & W3m3m2 & 13:58:22.4 & +52:43:19 & 2.56 & 2.93 & 0.31 & 2.3 & $23.9^{\pm 0.2}$ & $23.5^{\pm 0.1}$ & $23.2^{\pm 0.3}$ & $22.5^{\pm 0.2}$ & $0.55_{-0.10}^{+0.10}$ & 3 & - \\
arc37$_b$   & W3m3m2 & 13:58:46.6 & +52:21:00 & 1.58 & 2.25 & 0.00 & 1.1 & $26.2^{\pm 1.3}$ & $24.9^{\pm 0.4}$ & $24.2^{\pm 0.8}$ & $23.4^{\pm 0.3}$ & $0.35_{-0.09}^{+0.09}$ & 2 & - \\
arc38$_b$   & W4m0m1 & 22:12:29.9 & +00:17:25 & 2.47 & 2.80 & 0.00 & 2.4 & $24.6^{\pm 0.5}$ & $23.5^{\pm 0.1}$ & $23.1^{\pm 0.3}$ & $22.3^{\pm 0.1}$ & $0.58_{-0.10}^{+0.10}$ & 3 & - \\
arc39$_b$   & W4m0m2 & 22:13:07.1 & +00:30:37 & 3.06 & 3.75 & 0.32 & 2.5 & $24.9^{\pm 0.4}$ & $24.1^{\pm 0.1}$ & $23.3^{\pm 0.3}$ & $22.4^{\pm 0.1}$ & $0.58_{-0.10}^{+0.10}$ & 3 & SA122 \\
arc40$_b$   & W4p2m2 & 22:20:18.6 & +00:24:28 & 2.56 & 2.80 & 1.01 & 2.1 & $25.2^{\pm 0.6}$ & $24.4^{\pm 0.3}$ & $23.9^{\pm 0.6}$ & $23.0^{\pm 0.1}$ & $0.52_{-0.10}^{+0.10}$ & 2 & - \\
arc41$_b$   & W4p2m1 & 22:21:41.8 & +00:19:09 & 2.15 & 2.65 & 0.56 & 1.8 & $24.6^{\pm 0.3}$ & $23.4^{\pm 0.1}$ & $23.1^{\pm 0.2}$ & $22.7^{\pm 0.1}$ & $0.64_{-0.11}^{+0.11}$ & 2 & - \\
arc42$_c$   & W1m1p1 & 02:12:31.6 & -06:11:54 & 7.56 & 7.33 & 0.18 & 8.5 & $25.4^{\pm 0.5}$ & $23.9^{\pm 0.1}$ & $22.8^{\pm 0.2}$ & $21.8^{\pm 0.1}$ & $0.36_{-0.09}^{+0.09}$ & 3 & - \\
arc43$_c$   & W1m1p3 & 02:15:13.5 & -03:53:44 & 3.17 & 3.52 & 0.11 & 3.0 & $24.8^{\pm 0.5}$ & $24.1^{\pm 0.2}$ & $23.0^{\pm 0.2}$ & $22.1^{\pm 0.1}$ & $0.51_{-0.11}^{+0.10}$ & 2 & - \\
arc44$_c$   & W1m1p3 & 02:15:57.1 & -04:10:27 & 2.19 & 2.20 & 1.10 & 1.9 & $24.7^{\pm 0.5}$ & $24.1^{\pm 0.2}$ & $23.1^{\pm 0.2}$ & $22.4^{\pm 0.1}$ & $0.82_{-0.12}^{+0.12}$ & 3 & - \\
arc45$_c$   & W1m0p1 & 02:19:32.7 & -06:28:44 & 3.71 & 3.54 & 0.32 & 4.5 & $25.2^{\pm 0.5}$ & $24.5^{\pm 0.3}$ & $23.4^{\pm 0.2}$ & $22.4^{\pm 0.2}$ & $0.65_{-0.11}^{+0.11}$ & 2 & - \\
arc46$_c$   & W1p3p1 & 02:29:32.1 & -06:15:32 & 2.22 & 2.02 & 0.00 & 2.4 & $24.9^{\pm 0.4}$ & $24.2^{\pm 0.2}$ & $23.7^{\pm 0.3}$ & $22.8^{\pm 0.1}$ & $0.58_{-0.10}^{+0.10}$ & 2 & - \\
arc47$_c$   & W3m3m3 & 13:57:23.9 & +51:21:30 & 2.70 & 3.04 & 0.61 & 2.6 & $25.6^{\pm 0.7}$ & $24.3^{\pm 0.2}$ & $23.4^{\pm 0.3}$ & $22.3^{\pm 0.1}$ & $0.60_{-0.10}^{+0.11}$ & 2 & - \\
arc48$_c$   & W3m3m3 & 14:00:09.5 & +52:06:24 & 4.97 & 5.67 & 0.67 & 7.3 & $24.4^{\pm 0.3}$ & $23.2^{\pm 0.1}$ & $22.4^{\pm 0.1}$ & $21.5^{\pm 0.1}$ & $0.54_{-0.10}^{+0.10}$ & 2 & - \\
arc49$_c$   & W3m3m2 & 14:00:28.6 & +52:55:26 & 2.16 & 2.51 & 0.47 & 1.9 & $25.3^{\pm 0.6}$ & $24.3^{\pm 0.2}$ & $23.3^{\pm 0.3}$ & $22.6^{\pm 0.1}$ & $0.53_{-0.12}^{+0.10}$ & 2 & - \\
arc50$_c$   & W3m3m2 & 14:01:02.7 & +52:37:23 & 8.06 & 7.55 & 0.30 & 11.0 & $24.8^{\pm 0.4}$ & $23.3^{\pm 0.1}$ & $22.5^{\pm 0.2}$ & $21.6^{\pm 0.1}$ & $0.50_{-0.11}^{+0.10}$ & 3 & - \\
arc51$_c$   & W3m1m3 & 14:12:35.8 & +51:33:24 & 4.61 & 3.94 & 0.87 & 5.9 & $23.9^{\pm 0.3}$ & $22.8^{\pm 0.1}$ & $22.0^{\pm 0.1}$ & $21.2^{\pm 0.1}$ & $0.74_{-0.11}^{+0.11}$ & 3 & - \\
arc52$_c$   & W4m1m2 & 22:07:53.1 & +00:39:38 & 6.08 & 4.29 & 0.20 & 8.5 & $23.3^{\pm 0.2}$ & $22.6^{\pm 0.1}$ & $22.1^{\pm 0.1}$ & $21.4^{\pm 0.1}$ & $0.48_{-0.10}^{+0.10}$ & 2 & - \\
arc53$_c$   & W4m1m1 & 22:10:33.1 & +00:23:51 & 1.80 & 1.99 & 0.92 & 2.3 & $25.6^{\pm 0.7}$ & $24.4^{\pm 0.2}$ & $23.4^{\pm 0.5}$ & $22.4^{\pm 0.2}$ & $0.58_{-0.10}^{+0.10}$ & 2 & - \\
arc54$_c$   & W4p2m0 & 22:20:51.5 & +00:58:14 & 2.49 & 2.07 & 0.30 & 3.3 & $24.0^{\pm 0.2}$ & $23.3^{\pm 0.1}$ & $23.0^{\pm 0.2}$ & $22.5^{\pm 0.1}$ & $0.41_{-0.09}^{+0.09}$ & 2 & - \\
arc55$_c$   & W4p2m2 & 22:21:58.5 & +00:59:02 & 2.50 & 2.57 & 0.33 & 2.9 & $25.4^{\pm 0.7}$ & $24.2^{\pm 0.2}$ & $23.5^{\pm 0.7}$ & $22.8^{\pm 0.1}$ & $0.33_{-0.09}^{+0.09}$ & 2 & - \\
arc56$_c$   & W4p2m0 & 22:22:23.0 & +01:16:05 & 4.46 & 4.25 & 0.40 & 4.7 & $24.5^{\pm 0.6}$ & $23.1^{\pm 0.2}$ & $22.5^{\pm 0.3}$ & $21.9^{\pm 0.1}$ & $0.53_{-0.10}^{+0.10}$ & 2 & - \\
    \hline
  \end{tabular}
\end{sidewaystable*}

\begin{sidewaystable*}
  \vspace{18cm}  
  \centering
  \caption {Same as Table~(\ref{tab:catalog}) and (\ref{tab:catalog1})
    but referring to all detections not satisfying the colour
    criteria. This sample is mostly composed by candidates purely
    identified by visual inspection.}
  \label{tab:catalog2}
  \begin{tabular}{llrrrrrrrrrrrrr}
    ID & FIELD & RA & DEC & LEN & L/W & CUR & A & m$_{u\prime}$ & m$_{g\prime}$ & m$_{r\prime}$ & m$_{i\prime}$ & z$_{\rm lens}$ & RANK & CAT\\
    \hline
arc57$_a$   & W1m1p2 & 02:14:16.6 & -05:03:15 & 4.27 & 4.43 & 0.23 & 4.6 & $23.5^{\pm 0.2}$ & $22.8^{\pm 0.1}$ & $21.7^{\pm 0.1}$ & $21.0^{\pm 0.1}$ & $0.37_{-0.09}^{+0.09}$ & 3 & SL2SJ021416-050315 \\
arc58$_a$   & W1p1p1 & 02:22:33.6 & -06:06:24 & 3.33 & 5.41 & 0.00 & 2.3 & $24.6^{\pm 0.4}$ & $23.5^{\pm 0.1}$ & $22.7^{\pm 0.2}$ & $22.0^{\pm 0.1}$ & $0.45_{-0.11}^{+0.10}$ & 2 & - \\
arc59$_a$   & W1p2p3 & 02:24:01.0 & -03:46:27 & 4.90 & 7.20 & 0.24 & 6.3 & $24.6^{\pm 0.5}$ & $23.5^{\pm 0.2}$ & $23.1^{\pm 0.3}$ & $22.4^{\pm 0.3}$ & $0.81_{-0.12}^{+0.12}$ & 3 & SA42 \\
arc60$_a$   & W1p2p2 & 02:25:34.1 & -05:02:10 & 4.88 & 5.16 & 0.30 & 4.7 & $24.5^{\pm 0.4}$ & $23.9^{\pm 0.2}$ & $23.4^{\pm 0.4}$ & $22.6^{\pm 0.1}$ & $0.67_{-0.11}^{+0.11}$ & 3 & - \\
arc61$_a$   & W3m3m3 & 13:59:53.7 & +51:39:53 & 4.67 & 13.80 & 0.46 & 2.3 & $24.5^{\pm 0.3}$ & $23.4^{\pm 0.1}$ & $22.7^{\pm 0.2}$ & $22.0^{\pm 0.1}$ & $0.53_{-0.10}^{+0.10}$ & 2 & - \\
arc62$_b$   & W1p2p3 & 02:23:45.6 & -04:24:03 & 2.19 & 2.49 & 0.33 & 2.2 & $24.6^{\pm 0.3}$ & $23.1^{\pm 0.1}$ & $22.1^{\pm 0.1}$ & $21.7^{\pm 0.1}$ & $0.37_{-0.09}^{+0.09}$ & 2 & SL2SJ022345-042402 \\
arc63$_b$   & W1p3p2 & 02:28:34.3 & -05:09:48 & 1.50 & 2.66 & 0.01 & 1.0 & $26.2^{\pm 1.0}$ & $25.4^{\pm 0.5}$ & $24.9^{\pm 1.0}$ & $24.4^{\pm 0.8}$ & $0.57_{-0.10}^{+0.10}$ & 2 & - \\
arc64$_b$   & W1p3p1 & 02:30:11.6 & -05:50:21 & 3.29 & 3.16 & 0.05 & 3.8 & $23.7^{\pm 1.7}$ & $22.6^{\pm 0.2}$ & $0.0^{\pm 0.1}$ & $20.7^{\pm 0.1}$ & $0.49_{-0.21}^{+0.22}$ & 3 & SL2SJ023011-055023 \\
arc65$_c$   & W1m1p1 & 02:12:33.8 & -06:12:10 & 1.32 & 1.48 & 0.06 & 1.2 & $25.5^{\pm 0.5}$ & $24.1^{\pm 0.1}$ & $23.0^{\pm 0.2}$ & $22.3^{\pm 0.1}$ & $0.31_{-0.09}^{+0.11}$ & 2 & - \\
arc66$_c$   & W1m1p1 & 02:13:08.7 & -05:37:26 & 3.21 & 3.01 & 0.17 & 3.5 & $24.3^{\pm 0.2}$ & $22.7^{\pm 0.1}$ & $21.6^{\pm 0.1}$ & $21.0^{\pm 0.1}$ & $0.29_{-0.08}^{+0.08}$ & 2 & - \\
arc67$_c$   & W1m1p2 & 02:13:28.4 & -05:11:45 & 2.88 & 2.46 & 0.08 & 4.4 & $24.2^{\pm 0.3}$ & $23.1^{\pm 0.1}$ & $21.8^{\pm 0.1}$ & $21.0^{\pm 0.1}$ & $0.49_{-0.10}^{+0.10}$ & 2 & - \\
arc68$_c$   & W1m1p2 & 02:15:29.4 & -04:40:54 & 2.36 & 3.25 & 0.18 & 1.8 & $25.6^{\pm 1.1}$ & $24.2^{\pm 0.2}$ & $22.9^{\pm 0.2}$ & $22.1^{\pm 0.1}$ & $0.31_{-0.09}^{+0.09}$ & 3 & - \\
arc69$_c$   & W1m1p1 & 02:15:34.8 & -05:45:15 & 2.07 & 2.78 & 0.13 & 1.8 & $25.7^{\pm 0.8}$ & $23.9^{\pm 0.2}$ & $22.6^{\pm 0.2}$ & $22.0^{\pm 0.1}$ & $0.31_{-0.09}^{+0.09}$ & 2 & - \\
arc70$_c$   & W1m0p3 & 02:19:09.1 & -04:02:14 & 7.98 & 4.80 & 0.46 & 11.1 & $23.6^{\pm 0.2}$ & $22.5^{\pm 0.1}$ & $21.5^{\pm 0.1}$ & $20.8^{\pm 0.1}$ & $0.49_{-0.10}^{+0.10}$ & 3 & - \\
arc71$_c$   & W1p1p1 & 02:19:56.5 & -06:02:02 & 5.20 & 2.31 & 0.06 & 12.5 & $23.5^{\pm 0.2}$ & $22.5^{\pm 0.1}$ & $21.6^{\pm 0.1}$ & $20.5^{\pm 0.1}$ & $0.42_{-0.09}^{+0.09}$ & 2 & - \\
arc72$_c$   & W1p1m1 & 02:20:56.4 & -07:43:10 & 3.55 & 2.79 & 0.32 & 6.1 & $22.6^{\pm 0.2}$ & $22.1^{\pm 0.1}$ & $0.0^{\pm 0.0}$ & $20.8^{\pm 0.1}$ & $0.56_{-0.24}^{+0.23}$ & 3 & - \\
arc73$_c$   & W1p1p1 & 02:21:00.3 & -06:06:12 & 2.89 & 1.96 & 0.18 & 4.2 & $24.2^{\pm 0.4}$ & $22.3^{\pm 0.1}$ & $21.1^{\pm 0.1}$ & $20.5^{\pm 0.1}$ & $0.26_{-0.08}^{+0.10}$ & 2 & - \\
arc74$_c$   & W1p2p3 & 02:25:36.7 & -04:15:17 & 1.88 & 2.20 & 0.56 & 1.5 & $25.1^{\pm 0.4}$ & $24.1^{\pm 0.2}$ & $23.1^{\pm 0.1}$ & $22.1^{\pm 0.1}$ & $0.56_{-0.10}^{+0.10}$ & 2 & - \\
arc75$_c$   & W1p2p2 & 02:25:38.0 & -05:14:49 & 3.02 & 2.16 & 0.05 & 4.8 & $24.0^{\pm 0.2}$ & $22.7^{\pm 0.1}$ & $21.6^{\pm 0.1}$ & $21.0^{\pm 0.1}$ & $0.42_{-0.10}^{+0.09}$ & 2 & - \\
arc76$_c$   & W1p3p2 & 02:29:08.8 & -05:19:54 & 2.99 & 3.76 & 0.12 & 2.6 & $24.9^{\pm 0.5}$ & $23.8^{\pm 0.2}$ & $23.1^{\pm 0.2}$ & $22.4^{\pm 0.1}$ & $0.37_{-0.09}^{+0.09}$ & 2 & - \\
arc77$_c$   & W1p4m0 & 02:32:54.3 & -06:39:16 & 2.38 & 3.08 & 0.50 & 1.8 & $24.8^{\pm 0.4}$ & $24.2^{\pm 0.2}$ & $23.8^{\pm 0.5}$ & $23.5^{\pm 0.4}$ & $0.60_{-0.10}^{+0.11}$ & 2 & - \\
arc78$_c$   & W1p4m0 & 02:33:30.1 & -06:51:42 & 6.16 & 6.05 & 0.06 & 7.6 & $23.0^{\pm 0.2}$ & $22.4^{\pm 0.1}$ & $21.5^{\pm 0.1}$ & $20.5^{\pm 0.1}$ & $0.65_{-0.11}^{+0.11}$ & 1 & - \\
arc79$_c$   & W3m3m3 & 13:57:39.1 & +51:48:46 & 4.21 & 4.22 & 0.40 & 5.6 & $24.9^{\pm 0.4}$ & $23.6^{\pm 0.1}$ & $22.4^{\pm 0.2}$ & $21.1^{\pm 0.1}$ & $0.38_{-0.09}^{+0.09}$ & 3 & - \\
arc80$_c$   & W3m3m2 & 13:58:19.8 & +52:52:53 & 2.73 & 2.37 & 0.08 & 3.1 & $25.0^{\pm 0.5}$ & $23.3^{\pm 0.1}$ & $22.0^{\pm 0.1}$ & $21.2^{\pm 0.1}$ & $0.35_{-0.09}^{+0.09}$ & 2 & - \\
arc81$_c$   & W3m3m2 & 14:02:06.4 & +52:57:07 & 2.05 & 3.90 & 0.41 & 1.0 & $27.4^{\pm 3.6}$ & $24.9^{\pm 0.3}$ & $23.7^{\pm 0.4}$ & $22.5^{\pm 0.1}$ & $0.51_{-0.10}^{+0.10}$ & 2 & - \\
arc82$_c$   & W3m1m3 & 14:12:07.1 & +51:29:43 & 3.19 & 3.34 & 0.33 & 3.9 & $25.3^{\pm 0.7}$ & $23.6^{\pm 0.1}$ & $22.5^{\pm 0.1}$ & $21.1^{\pm 0.1}$ & $0.30_{-0.09}^{+0.13}$ & 2 & - \\
arc83$_c$   & W4m1m1 & 22:08:51.3 & +00:46:22 & 1.83 & 2.61 & 0.50 & 1.4 & $24.9^{\pm 0.4}$ & $23.9^{\pm 0.1}$ & $22.8^{\pm 0.2}$ & $22.0^{\pm 0.1}$ & $0.49_{-0.21}^{+0.22}$ & 2 & - \\
arc84$_c$   & W4m1m2 & 22:09:35.5 & +00:31:26 & 9.09 & 6.08 & 0.16 & 13.0 & $24.0^{\pm 0.2}$ & $22.2^{\pm 0.1}$ & $21.4^{\pm 0.1}$ & $21.0^{\pm 0.1}$ & $0.32_{-0.09}^{+0.09}$ & 2 & - \\
arc85$_c$   & W4m0m1 & 22:12:35.5 & +00:43:11 & 3.75 & 2.92 & 0.77 & 4.7 & $23.9^{\pm 0.2}$ & $22.9^{\pm 0.1}$ & $22.0^{\pm 0.1}$ & $21.6^{\pm 0.1}$ & $0.29_{-0.08}^{+0.11}$ & 2 & - \\
arc86$_c$   & W4m0m2 & 22:13:28.9 & +00:44:53 & 5.99 & 8.24 & 0.42 & 4.8 & $24.7^{\pm 0.3}$ & $23.4^{\pm 0.1}$ & $22.1^{\pm 0.1}$ & $21.1^{\pm 0.1}$ & $0.38_{-0.09}^{+0.09}$ & 3 & - \\
arc87$_c$   & W4m0m1 & 22:15:02.6 & +00:49:50 & 3.74 & 3.69 & 0.53 & 3.8 & $25.0^{\pm 0.4}$ & $23.7^{\pm 0.1}$ & $22.3^{\pm 0.1}$ & $21.4^{\pm 0.1}$ & $0.51_{-0.10}^{+0.10}$ & 3 & - \\
arc88$_c$   & W4p1m1 & 22:18:21.1 & +00:01:53 & 2.00 & 1.60 & 0.58 & 2.8 & $23.9^{\pm 0.2}$ & $23.1^{\pm 0.1}$ & $22.4^{\pm 0.1}$ & $21.7^{\pm 0.1}$ & $0.50_{-0.10}^{+0.10}$ & 3 & - \\
arc89$_c$   & W4p2m2 & 22:19:00.9 & +00:54:56 & 2.58 & 2.24 & 0.19 & 2.7 & $24.9^{\pm 0.8}$ & $22.5^{\pm 0.2}$ & $21.3^{\pm 0.3}$ & $20.4^{\pm 0.1}$ & $0.46_{-0.10}^{+0.10}$ & 2 & - \\
arc90$_c$   & W4p2m1 & 22:22:17.6 & +00:12:02 & 2.19 & 2.57 & 0.45 & 2.1 & $24.9^{\pm 0.4}$ & $23.5^{\pm 0.1}$ & $22.5^{\pm 0.1}$ & $21.7^{\pm 0.1}$ & $0.51_{-0.10}^{+0.10}$ & 3 & - \\
    \hline
  \end{tabular}
\end{sidewaystable*}

\begin{table}[!]
  \caption {Completeness of the final arc candidates sample listed in
    Tables~(\ref{tab:catalog}, \ref{tab:catalog1} and
    \ref{tab:catalog2}), comparing the results obtained before and
    after the colour selection. Columns refer to the objects detected
    by the arcfinder (arcf., labelled with `a'), those segmented but
    not recognized as arcs by the arcfinder (arcf. seg., labelled with
    `b'), those identified with the visual inspection only (visual
    only, labelled with `c' in the sample catalogue) and the complete
    sample (all). This demonstrates how the colour selection looses
    only 17\% of the single-band arcfinder sample (first column). The
    loss is greater when the visual detections are included, 48\%,
    since many of those have poor photometry, being blended, faint or
    fragmented (last column).}
  \centering
  \label{tab:completeness}
  \begin{tabular}{r|ccc|c}
    & arcf. (a) & arcf. seg. (b) & visual only (c) & all \\
    \hline
    before col. sel. & 29 & 20 & 41 & 90 \\
    after col. sel. & 24 & 17 & 15 & 56 \\
    \hline
    completeness & 83\% & 85\% & 36\% & 62\%\\
  \end{tabular}
\end{table}

\section{Completeness and purity of the colour-selected sample}\label{sec:complete}

We now evaluate the completeness and contamination of the sample,
discussed in Section~(\ref{sec:visual}) for these three candidate
subsamples:
\begin{enumerate}[a)]
  \item 29 objects, among 36 026 detected by the arcfinder
  \item 20 objects, among 165 673 segmented but discarded by the arcfinder's
    morphological filter,
  \item 41 objects identified only with the visual inspection,
\end{enumerate}
and subdivided in those passing the colour filtering and those which
do not, both within their subsamples and with respect to the total
sample. To evaluate the sample completeness, only detections that
passed visual inspection have to be used. In contrast, the complete
sample has to be used for the evaluation of the contamination level.

The 20 objects segmented by the arcfinder but failing to fulfil all
geometrical constraints were the results of the initial calibration
run with reduced filtering, not the final detection process. Since
they were rejected, counting them as arcfinder detections would be
misleading.  Since they illustrate the impact of the morphological
filter, we still keep them as separate items.  The full automatic
procedure for the selection of candidates is represented by the 29
objects recognized by the arcfinder and passing all of its
filters. While this number might seem to be small, we have to keep in
mind that it is comparable to the number of arc candidates identified
in the CARS fields by previous works and detected with the combined
use of arcfinders and visual inspection
\citep[e.g.][]{2012ApJ...749...38M}.

The completeness of the colour selection for the three subsamples and
of the total sample is detailed in Table~(\ref{tab:completeness}). As
we show there, the multi-band arcfinder implementing the colour
selection is complete at a 83\% level with respect to the single-band
arcfinder for our sample. Very similar results are obtained for the
objects segmented but in a further filtering step discarded by the
arcfinder due to their geometrical properties: here, the completeness
of the colour sample is at a 85\% level. The completeness is smaller,
at 62\%, once the detections idsentified only by visual inspection are
included in the complete sample. This is because these detections are
faint, fragmented or heavily blended and therefore have very poor
photometry and consequently not well defined colours. This is clear
from Figure~(\ref{fig:gr-ri_flux}) where these objects fall in a
colour-colour region where no other fall.

Finally, the contamination level is measured by evaluating the ratio
between the number of sources detected by the arcfinder but considered
not to be arcs before ($35\,997$ objects) and after ($5\,573$ objects)
the colour selection. The improvement gained by the use of colour
information is evident: a reduction of a factor of $6.5$ in the sample
contamination. This result largely compensates the relatively small
loss of detections due to multi-band filtering. Even if the sample is
still heavily contaminated and other criteria have do be included to
move in the direction of a fully automated method, such a strong
reduction in the sample contamination is of crucial importance for the
present and for upcoming surveys, which will cover between $1\,500\;deg^2$
and $15\,000\;deg^2$ in the near future \citep[see e.g. KiDS and
  EUCLID,][]{2012ExA...tmp...34D,2011arXiv1110.3193L} and which can
only be handled in an automatic or semi-automatic way.

\section{Conclusions}

The present and upcoming wide field surveys, such as for example the
KiDS survey with $1\,500$ square degrees or the ESA Euclid mission
with $15\,000$ square degrees of sky coverage, are posing a pressing
need for a method to automatically detect strong lensing features in a
reliable way. The minimization of human intervention in the detection
process is fundamental to avoid a massive use of eye-ball checking,
which is subjective and extremely time consuming. We proposed a
detection method and tested it against the CARS data employing, on one
hand, a tailored image segmentation based on coherent patterns in
local second brightness moments, which is capable of isolating
elongated objects and of retrieving their length and width,
fundamental quantities to characterize and select the gravitational
arc candidates \citep{2007A&A...472..341S}. On the other hand, a
colour-colour based selection, in our case $g^\prime-r^\prime$ against
$r^\prime-i^\prime$, is motivated by the fact that most of the lensed
sources have similar emission properties. We describe this behaviour
with a model reproducing the expected redshift distribution of
arcs. The model implies the lensing optical depth expected in a
$\Lambda$CDM cosmology which, in combination with the sources'
redshift distribution of the CARS galaxy catalogues, peaks for sources
at redshift $z\sim 1$. By further modelling the colours derived from
the SED of the galaxies dominating the population at that redshift,
the model well reproduces the colour of the observed arcs. The colour
selection of arcs is not a limitation, because the population of the
lensed sources selected in this way is the most numerous, ensuring
that only a small fraction of arcs is lost. It does not bias measures
derived from gravitational lensing, which is a completely achromatic
phenomenon and the sources are not correlated with the lenses because
of their large separation along the line of sight.

To verify the reliability and efficiency of the colour selection we
applied our procedure to the CARS data consisting in $37\;deg^2$
derived from the $W1$, $W3$ and $W4$ fields of CFHTLS, and containing
$7\; 10^6$ (detected with SExtractor). Each step can be summarized as
follows:
\begin{enumerate}
\item {\it Preprocessing}: we create a weighted co-addition of the
  $g^\prime$ and $r^\prime$ stacks to produce an image with an
  enhanced signal-to-noise ratio, over which we run the arcfinder. The
  $u^\star$ and $z^\prime$ bands are ignored because of their low S/N
  ratio, as well as the $i^\prime$ band to avoid blending, potentially
  caused by the lens (likely one or more massive elliptical galaxies)
  being in the vicinity of a gravitational arc, which we expect to be
  more severe in this band.

\item {\it Arcfinder segmentation}: we segment images with the
  arc\-finder to detect all elongated objects and derive their
  geometrical properties, such as length, length-to-width ratio and
  curvature. These geometrical properties are used to select the
  objects with $L>2.9$ arcsec (15 pixels) and $L/W_{min}>4$, obtaining
  a sample of $36\,026$ candidates ($\sim 970$ per square degree).

\item {\it Colour selection}: the largest probability of obtaining
  strongly elongated arcs is given by sources located at redshift
  $z\approx 1$, where the population is dominated by galaxies with
  large star forming regions characterized by relatively uniform colour
  properties. We thus selected the sources with respect to their
  ``colour-colour'' (${\rm f_g-f_r,f_r-f_i}$) diagram shown in
  Figure~(\ref{fig:gr-ri_flux}). This returns a sample of $5\,597$
  candidates automatically selected.

\end{enumerate}
Note that data acquired by space-based instruments would return
results with more depth and, more importantly, purity, not only
because of larger sensitivity but also because of the higher
resolution. In fact, because of their very small width gravitational
arcs are very sensitive to the image PSF, easily becoming shallower
and of smaller $L/W$ ratio, which is the most stringent property to
discriminate them from other astrophysical sources.

To validate the candidates we visually inspected their colour
composite images (based on the $g^\prime$, $r^\prime$ and $i^\prime$
bands) removing clear spiral galaxies and image ``artefacts'' caused
mainly by bright stars and blended sources. We selected only the arcs
with a curvature compatible with a possible lens such as a large
elliptical or an evident over-density of galaxies.  This led to the
final sample of 49 arc candidates, 20 of which were segmented by the
arcfinder but did not satisfy the geometrical constraints we imposed
to define arcs and were therefore not recognized as such by the
automatic procedure. In addition, we visually inspected the full
survey in order to evaluate the completeness and purity of the sample
retrieved by the arcfinder against such an inspection.  In this
process, an additional sample of 41 detections, found purely by visual
inspection, were included in the final catalogue, which in the end
lists 90 objects, 73 of which new proposed arc candidates.

In conclusion, a fully automatic arcfinder is not jet available but
the colour selection discussed in the paper turned out to be very
effective in isolating the most promising arc candidates: while
completeness is decreased to 83\% with respect to the single-band
arcfinder, the sample contamination is drastically reduced by a factor
of 6.5. This large gain in purity is of crucial importance for the
existing and upcoming surveys, which will cover thousands of square
degrees and can only be handled in an automatic or semi-automatic way.

\begin{acknowledgements}
  This work was supported by the Transregional Collaborative Research
  Centre TRR~33 (MM). S.M. is supported by contract research
  "Internationale Spitzenforschung II/2-6" of the Baden W\"urttemberg
  Stiftung.
\end{acknowledgements}

\appendix

\section{Arcfinder segmentation}\label{app:arcfinder}

We briefly describe the arcfinder method introduced in
\cite{2007A&A...472..341S}, the automatic filtering, and
post-processing that was added to the basic detection algorithm. This
segmentation and first filtering was used for the final colour
selection discussed in the paper.

\subsection*{Introduction}

To measure the orientation of a local source pattern, e.g. an area of
higher flux inside an ellipse, a simple formalism can be used: the
second moments
\begin{equation}\label{eq:Qij}
  Q_{ij} \propto \int_{A}(x_i-\bar{x}_i)(x_j-\bar{x}_j)I(\vec{x})\,\mathrm{d}^2x
\end{equation}
in an area $A$ surrounding the pattern at $(\bar{x}_1,\bar{x}_2)$ and
- usually reweighed - pixel values $I$ combine into an ellipticity
vector $(Q_{11}-Q_{22},2Q_{12})^T$ that encloses twice the angle to
the $x_1$ axis as the pattern itself. For an ellipse, we recover its
orientation as long as the area centre is on the major axis. On a
constant slope the orientation is parallel to the gradient vector. For
an extended source, like a gravitational lensing arc, we can derive
its local orientation as long as the area is (1) large enough to
compensate for noise and (2) small enough to avoid blending with other
sources and effects from any arc curvature. Also, to recover the
orientation directly, the centre pixels must be selected, such that
they are close to the top of the feature and the integration area is
not measuring the slope on both sides.

\subsection*{Arc detection}
\begin{figure}
  \centering
  \includegraphics[width=0.488\hsize]{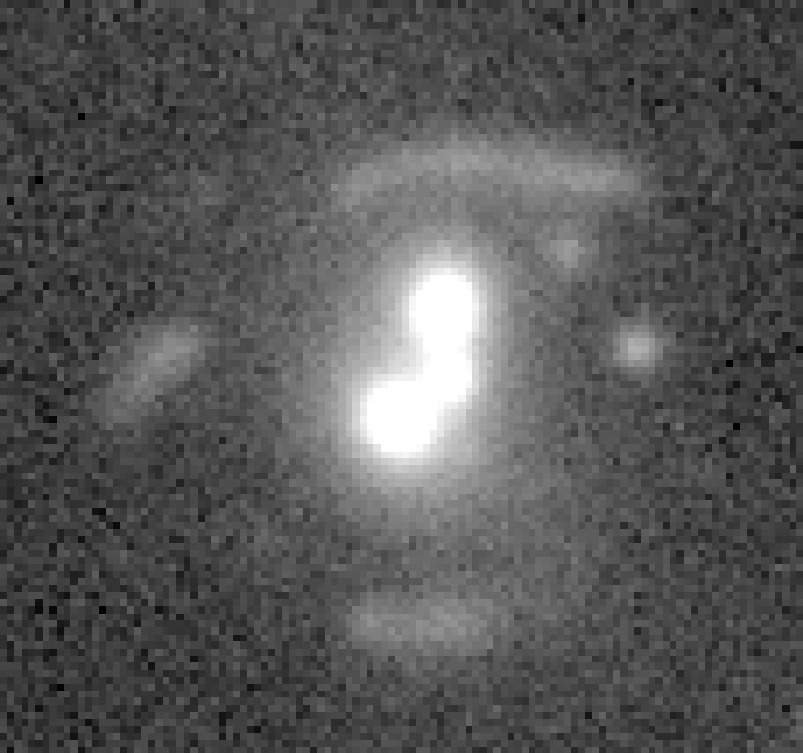}
  \includegraphics[width=0.488\hsize]{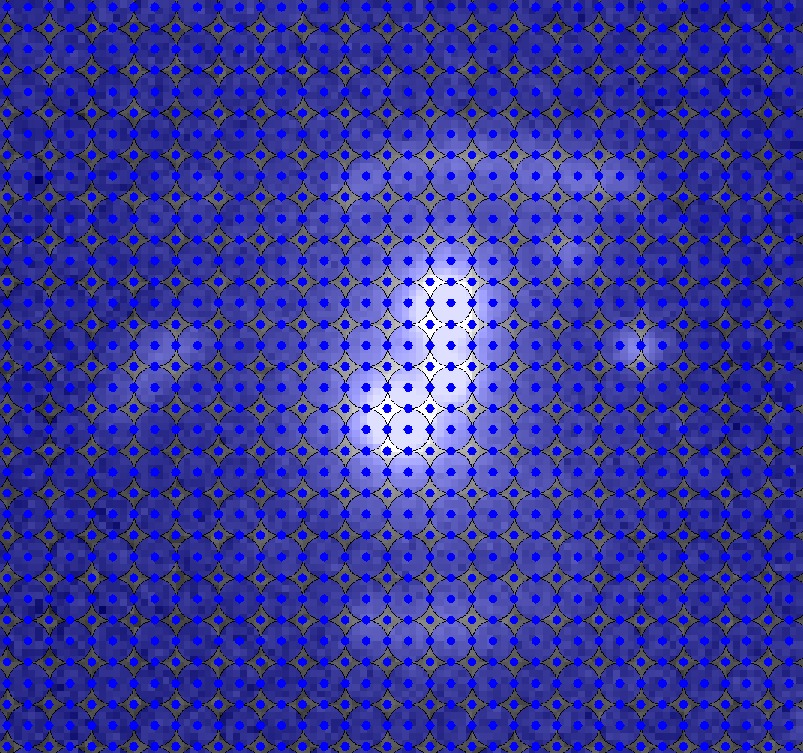}\\
  \includegraphics[width=0.488\hsize]{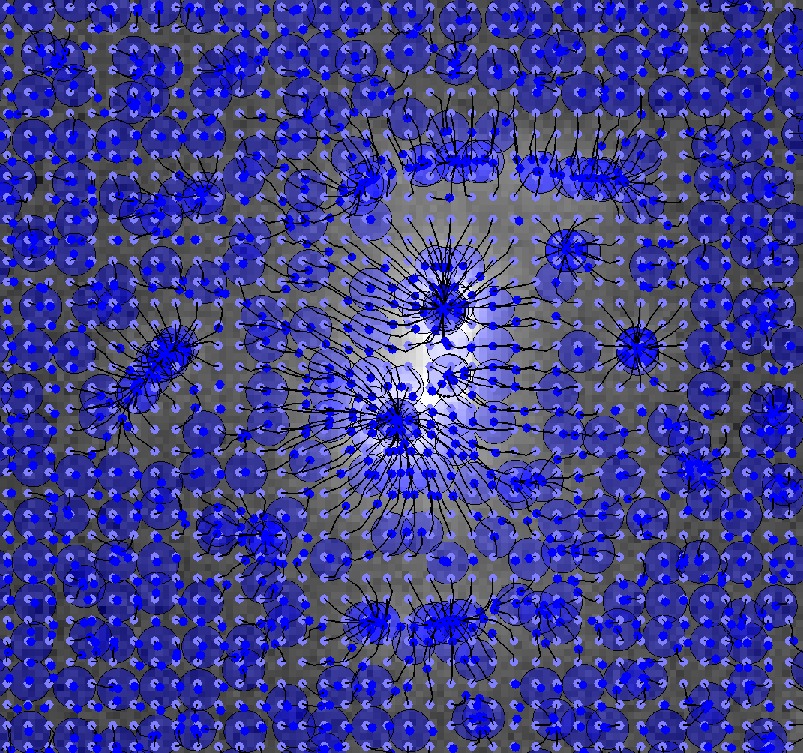}
  \includegraphics[width=0.488\hsize]{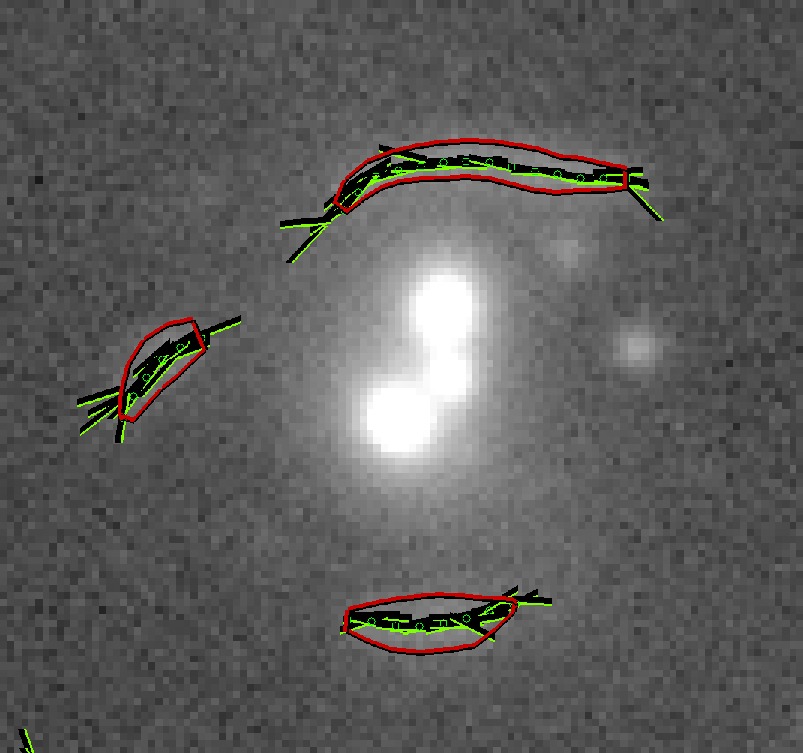}
\caption{Top-left: I-band SL2S J02140-0535 observation, also featuring
  a giant arc. Top-right: regular grid of disk shaped areas providing
  the initial condition of the displacement iterations. Only every
  second row and column of areas are shown. Bottom-left: areas after
  three iterations, each moving them to their centre of brightness,
  and the path they traced. Bottom-right: local orientations for each
  area, determined using the second brightness moments. Lines
  corresponding to coherent areas are highlighted. The red contours
  show the segmentation of the objects.}
\label{figure:a2390_814_detection}
\end{figure}
The arcfinder method tests for coherent orientations in areas
preferentially centred on features of locally higher intensity. To
achieve this, a regular grid of overlapping, disk shaped areas of
equal radii is first placed on the complete image. In three
iterations, each area is then displaced to its previous centre of
brightness, using a reweighed pixel flux. Areas originally close to a
feature move up the slope on both sides, increasing their number
density on any significant features' ridge-lines. On a flat or
constantly sloped background, areas only move relative to each other
due to pixel noise. At their final positions, each area's orientation
is measured using second moments as described above. To determine
coherence, the relative orientations together with the relative
placement of areas initially placed close to each other are taken into
account. Finally, the centre coordinates of coherent areas are grouped
together using a friend-of-friend type algorithm. These groups of
coordinates correspond to the initial arcfinder detections.

\subsection*{Filters and post processing}

To reduce the number of spurious detections and to expand on the
spatial information provided by the area positions and orientations in
each detection, filters are applied to the original detections and the
shape and flux of each detection are determined. Each of the small
detection areas is subjected to a simple fitting procedure that
preferentially removes areas centred on point sources and on a
background of constant noise without significant structure in the flux
distribution. Then, detections with less than a minimal amount of
valid areas left or shorter than a threshold are removed. Using a
fully automatic active contour evolution method \citep{Kass88},
isophote contours around each detection are determined, which in turn
allows for the determination of the flux. Using the flux and a
measurement of the background noise in the larger area surrounding
each detection, the signal to noise ratio is set. A more precise
length and a length-to-width ratio is computed from the contour, and
minimal thresholds are applied to both photometric and morphological
data.

\begin{table*}[b]
  \label{tab:parameters}
  \vspace{10cm}
  \caption{List of the arcfinder parameters used in this work. For a
    detailed description refer to the text.}
  \begin{center}
    \begin{tabular}{lll}
      \hline
      Ridge detection & & \\
      \hline
      Gridsize & 7 & Length and width of each cell [pix]\\
      Cell spacing & 0.45 & Distance between cells is Cell spacing x
      Gridsize\\
      Threshold cell & 0.45 & Coupling threshold for average coupling
      over the cell neighbourhood\\
      Threshold object & 0.75 & Coupling threshold for inclusion of a
      single cell into an object\\
      Threshold graph & 0.75 & Coupling threshold for graph generation\\
      \hline
      Filtering & &\\
      \hline
      Deblending asymmetry & 30 & median intensity difference for deblending\\
      Deblending distance & 21 & minimum deblending distance\\
      Stars saturation intensity & 500 & critical intensity above which pixels are considered as saturated\\
      Star intensity & 50 & intensity above which pixels are likely to belong to stars creating diffraction spikes\\
      Star-fluxcoeff spike & 3200 & compute spike radius as fluxcoeff-spike$\;\times\;{\rm flux}^{1/2}$\\
      Star-fluxcoeff disk & 420 & compute disk radius as fluxcoeff-disk$\;\times\; {\rm flux}^{1/3}$\\
      Maxpeakflux & 0.6 & maximal peak flux in ADUs above the background\\
      \hline
    \end{tabular}
  \end{center}
\end{table*}

\end{document}